\documentclass[a4paper,fleqn]{cas-sc}

\usepackage[authoryear,longnamesfirst]{natbib}
\usepackage{bm}
\usepackage{bigints}
\usepackage{amsmath}
\newcommand{\inv}[1]{#1^{-1}}
\usepackage{enumitem}
\usepackage{float}
\usepackage{placeins}
\usepackage{graphicx}
\usepackage{subcaption}
\usepackage{pdflscape}

\def\tsc#1{\csdef{#1}{\textsc{\lowercase{#1}}\xspace}}
\tsc{WGM}
\tsc{QE}

\begin{document}
\let\WriteBookmarks\relax
\def\floatpagepagefraction{1}
\def\textpagefraction{.001}

\shorttitle{}    

\shortauthors{}  

\title [mode = title]{A spatio-temporal block aggregation model for latent log Gaussian outcomes: application on modelling wastewater virus concentration in Wales}  



%
\author[1]{Stephen Jun Villejo}[orcid=0000-0002-0510-3143]
\ead{s.villejo@imperial.ac.uk}
\credit{}

\cormark[1]
\author[2]{Peter Diggle}[orcid=0000-0003-3521-5020]
\credit{}
\author[3]{Guangquan Li}[orcid=0000-0002-8736-5349]
\credit{}
\author[1]{Ella White}[orcid=0000-0002-0510-3143]
\credit{}
\author[4]{Matthew Wade}[orcid=0000-0001-9824-7121]
\credit{}
\author[5]{Christopher Williams}[orcid=0000-0002-5092-4987]
\credit{}
\author[6]{Davey L. Jones}[orcid=0000-0002-1482-4209]
\credit{}
\author[5]{Alisha Davies}[orcid=0000-0002-8066-7264]
\credit{}
\author[1]{Marta Blangiardo}[orcid=0000-0002-1621-704X]

\credit{}

\affiliation[1]{organization={Faculty of Medicine, Imperial College London},
            city={London},
            postcode={W12 0BZ}, 
            country={UK}}

\affiliation[2]{organization={Faculty of Health and Medicine, Lancaster University},
            postcode={LA1 4AT}, 
            country={UK}}
\affiliation[3]{organization={School of Engineering, Physics and Mathematics, Northumbria University},
            postcode={NE1 8SU}, 
            country={UK}}
\affiliation[4]{UK Health Security Agency, E14 4PU, UK}
\affiliation[5]{organization = {Public Health Wales},
city = {Cardiff},
postcode = {CF10 4BZ},
country = {UK}}
\affiliation[6]{organization={Environment Centre Wales, Bangor University},
  city={Bangor},
  postcode={LL57 2UW},
  country={UK}}
\cortext[1]{Corresponding author}

\begin{abstract}
Wastewater-based epidemiology has emerged as a valuable tool for monitoring community-level infectious disease dynamics, providing population-wide signals that complement clinical surveillance. However, wastewater measurements are often observed as aggregated values over irregular  spatial units. This work develops an approach to link an underlying spatially continuous processes and an aggregated outcome.
We propose a spatio-temporal model for latent log-Gaussian outcomes that provides a coherent framework for inference and prediction, allowing the process to be integrated over arbitrary spatial configurations. This framework can also be used for subsequent analyses, such as linking wastewater signal to health outcomes at administrative areas. We use a Bayesian framework for inference via the linearised integrated nested Laplace approximation (INLA) approach. We apply the proposed methodology to model SARS-CoV-2 N1 gene copies in wastewater across 47 catchment areas in Wales from the beginning of August 2022 to the end of July 2023. The results demonstrate that the model captures spatial and temporal patterns and has good predictive performance. Results also show that estimated viral gene copies are strongly linked to positivity rates from COVID-19 PCR tests at the local authority level. Our findings highlight the importance of explicitly modelling block aggregation when analysing wastewater surveillance data. The proposed framework provides a flexible and principled approach for integrating environmental surveillance data into public health monitoring systems.

\end{abstract}


\maketitle

\section{Introduction}\label{sec:intro}

In many applied settings, observed data arise as spatially aggregated summaries of an underlying spatially continuous process \citep{diggle2013spatial} due to confidentiality issues or limitations of the data collection process. Since covariates and prediction targets are often defined at different, typically finer spatial resolutions than the aggregated outcome, this mismatch in spatial support, referred to as spatial misalignment, introduces a fundamental inferential challenge.  A crude solution to this problem is to aggregate or `re-align' the covariate data to match the outcome (response) data resolution. However, this approach proves inferior compared to explicitly modelling the latent spatially continuous process and keeping this distinct from the sampling model, which we achieve through a class of models referred to as \textit{block aggregation models} and recently proposed by \cite{villejo2026spatially}. This modelling approach provides a coherent framework for parameter inference, predictions at new areal configurations, and spatial disaggregation.  Related work is  discussed in \cite{suen2026coherent} and in \cite{rutten2026bayesian}.

A particularly compelling setting in which spatial misalignment arises naturally is wastewater-based epidemiology (WBE). By analysing viral biomarkers shed in human waste and transported through sewer networks, WBE provides a population-level signal that reflects infection levels across entire catchment populations. WBE has emerged as a powerful surveillance tool for monitoring infectious diseases \citep{li2024integrating, mills2024utility, li2023spatio, weidhaas2021correlation}. During the COVID-19 pandemic, WBE was rapidly adopted worldwide as a complementary surveillance system and  proved able to detect early increases in SARS-CoV-2 transmission, and to provide estimates of disease prevalence when clinical testing was limited or scaled back \citep{manuel2022strengthening, wade2022understanding, medema2020presence}. However, a fundamental challenge in WBE is that observations are collected over infrastructure-defined catchment areas, which rarely align with the administrative boundaries at which public health decisions are made and epidemiological data are reported. Several approaches have been proposed to address this misalignment, although most treat it deterministically rather than probabilistically. \cite{li2023spatio} proposed a spatio-temporal Bayesian geostatistical model for SARS-CoV-2 concentrations at sewage treatment works in England, predicting virus concentration at fine spatial scales and subsequently aggregating to local authority level using population weighting. \cite{mills2024utility} and \cite{morvan2022analysis} adopted deterministic spatial mapping approaches, re-expressing wastewater concentrations at administrative geographies through area-weighted intersections between catchment areas and target regions prior to modelling. These approaches do not propagate the uncertainty arising from the spatial aggregation process into downstream inference.

 The misalignment between the spatial support of wastewater measurements and of health outcome data makes WBE a natural and important application domain for block aggregation models. 
 However, the existing framework proposed in \cite{villejo2026spatially} is not directly suited to log-Gaussian outcomes observed as aggregated values over irregular areas, as its implementation requires the sampling model to have a distributional form that is closed under aggregation. 
  In this work, we extend the framework of \cite{villejo2026spatially} to the spatio-temporal setting and propose a spatio-temporal block aggregation model for latent log-Gaussian outcomes observed as aggregated values over irregular areal units. The model assumes that the underlying process lives in a spatially continuous domain and explicitly embeds the aggregation mechanism within the data likelihood, distinguishing it from approaches that treat spatial re-alignment as a pre-processing or post-processing step. The distributional assumption is particularly relevant when the distribution of the areal outcome is highly skewed  and covariate information, such as population size and land cover characteristics, is available
at a finer, nested spatial resolution. In contrast to \cite{li2023spatio}, who specify the predictor expression at the catchment level and aggregate predictions post-hoc using population weighting, our model accounts for the fact that each observed catchment-level measurement is an aggregation of latent unobserved values from contributing nested areas and propagates this uncertainty coherently. The proposed framework supports predictions at arbitrary spatial configurations, enabling seamless aggregation to any target areal partition, such as Lower Super Output Areas (LSOAs) or Lower Tier Local Authorities (LTLAs), without requiring model refitting.
  
We apply the proposed methodology to model the viral load of SARS-CoV-2 (N1 gene) in wastewater across 47 catchment areas in Wales, covering approximately 80\% of the Welsh population, from August 2022 to July 2023. The primary objective is to predict viral load at the level of the 22 LTLAs in Wales, which are the administrative regions responsible for collecting official health-related data. We use a Bayesian inferential framework via the linearised integrated nested Laplace approximation (INLA) \citep{rue2009approximate}, implemented through the \texttt{inlabru} library \citep{lindgren2024inlabru}. We also present exploratory results on the association between the predicted wastewater signal and COVID-19 polymerase chain reaction (PCR) positivity rates at the LTLA level, demonstrating the potential of the proposed framework  for integration of environmental surveillance data into public health monitoring systems. A more extensive integration model with clinical health outcomes is left as future work.

The remainder of the paper is structured as follows. Section \ref{sec:model} discusses the model, while Section \ref{sec:inference} discusses the Bayesian inferential algorithm.
We present an application to wastewater virus concentration in Section \ref{sec:application}, give results in Section \ref{sec:results}, and end with a discussion in Section \ref{sec:discussion}.

\section{Model}\label{sec:model}

\subsection{Latent process and fine-scale model}\label{fine-scale}

Let $\{S(x,t) : x \in D \subset  \mathbb{R}^2, t=1,\ldots, T\}$ denote the underlying spatially continuous process of interest, defined over a spatial domain $D$ and observed at discrete time points. In practice we do not observe $S(x,t)$ directly; instead, we observe aggregated summaries over a set of irregular areal units. To model this, we introduce fine-scale areal units at which the latent process is defined and covariate information is available.

Specifically, let $C_i, i=1,\ldots,N$ denote the areal units (e.g. catchment areas) over which aggregated observations are collected and let   $\{b_{ij}: j=1,\ldots,J_i\}$ denote the fine-scale areal units.  We define the latent model at the level of these fine-scale areal units. Defining $Z_{ijt}$ as the unobserved quantity of interest (here the number of gene copies) for each $b_{ij}$ and time $t$, we specify: 
\[Z_{ijt} \sim \log \text{Normal}\big(\mu_{ijt},\sigma^2_Z\big), \;\;\; j=1,\ldots,J_i; \; t=1,\ldots,T\]
where $\mu_{ijt}$ is the mean of the log-transformed latent quantity, modelled as:
\[\mu_{ijt} = \beta_0 + \bm{\beta}^{\intercal} \bm{x}_{ijt} + \xi_{ijt}\]

Here, $\mu_{ijt}$ is the spatially continuous process, $\bm{x}_{ijt}$ are covariates, $\beta_0$ and $\bm{\beta}$ are fixed effects, and $\xi_{ijt}$ is a stationary, zero-mean Gaussian process.

\subsection{Aggregation and sampling model}\label{Aggregation}

In the WBE context, the total number of gene copies measured on catchment $C_i$ at time $t$ is the sum of contributions from all fine-scale areal units: 
\[S_{it}=\sum_{j=1}^{J_i} Z_{ijt}\] 
This aggregation is physically motivated: the viral load measured at a sewage treatment works reflects the cumulative shedding from all contributing sub-populations within the catchment. The first two moments of $S_{it}$ follow from standard properties of the log Gaussian distribution:
\begin{equation}\label{eq:first_second_moments_Sit}
    \begin{aligned}
        &\mathbb{E}[S_{it}] = \sum_{j=1}^{J_i} \mathbb{E}\Big[Z_{ijt}\Big] = \sum_{j=1}^{J_i} \exp\big\{\mu_{ijt} + \tfrac{1}{2}\sigma^2_Z\big\} \\
        &\text{Cov}[S_{it},S_{i^*\ell}] = \sum_{j=1}^{J_i} \sum_{k=1}^{J_{i^*}}\text{Cov}[Z_{ijt},Z_{i^*k\ell }]; \qquad i^*=1,\ldots,N; \qquad \ell=1,\ldots, T\\
        &\;\;\;\;\;\;\;\;\;\;\;\;\;\;\;\;\;\;\;\;= \sum_{j=1}^{J_i}\sum_{k=1}^{J_{i^*}}\exp\big\{ \mu_{ijt}+\mu_{i^*k\ell} + \sigma^2_Z \big\}\times\Big[\exp\big\{\text{Cov}(\log Z_{ijt},\log Z_{i^*k\ell})\big\}-1\Big].
        \end{aligned}
\end{equation}

Since the $Z_{ijt}$ are correlated log-Gaussian random variables, their sum $S_{it}$ does not, in general, admit a closed-form distribution. However, empirical evidence of strong right-skewness in wastewater measurements motivates approximating $S_{it}$ as log Gaussian:
\[S_{it} \approx \log \mathrm{Normal}(\mu_{S_{it}},\sigma^2_{S_{it}}).\]
This is a moment-matching approximation: the parameter $\mu_{S_{it}}$ is chosen so that the first moment of the approximating log Gaussian distribution matches the first moment of the true distribution of $S_{it}$. The derivation is given in Appendix A; here, we summarise the key results. We have the following moment estimators:
\begin{equation}\label{eq:logsumexp}
\begin{aligned}
&\mu_{S_{it}} = \log \left( \sum_{j=1}^{J_i} \exp\Big\{\mu_{ijt}+\dfrac{1}{2}\sigma^2_Z\Big\} \right) - \frac{1}{2}\sigma^2_{S_{it}} \\
&\sigma^2_{S_{it}} = \log \Bigg( \dfrac{\sum_{j=1}^{J_i} \Big(\exp\big\{\sigma^2_Z\big\}-1\Big)\Big(\exp\big\{ 2\mu_{ijt} + \sigma^2_Z\big\}\Big)  +  2\sum_{j<k}\text{Cov}(Z_{ijt},Z_{ikt})}{\Big( \sum_{j=1}^{J_i}\exp\Big\{\mu_{ijt}+\dfrac{1}{2}\sigma^2_Z\Big\} \Big)^2} + 1\Bigg), 
\end{aligned}
\end{equation}
where $\text{Cov}(Z_{ijt},Z_{ikt}) =\exp\big\{ \mu_{ijt} + \mu_{ikt} + \sigma^2_Z \big\}\times\Big[\exp\big\{c_{ijk,t}\big\}-1\Big]$, with $c_{ijk,t} = \text{Cov}(\log Z_{ijt},\log Z_{ikt})$.

The approximation is expected to be most accurate when $\sigma^2_Z$ is small, in which case the log Gaussian components are tightly concentrated and the sum is well-approximated by a log Gaussian distribution via moment matching. Also, note that the expression for $\sigma^2_{S_{it}}$ is the log of a squared coefficient of variation. When the latent $\mathbb{E}[Z_{ijt}]$ are large and $\sigma^2_Z$ is small, then $\sigma^2_{S_{it}}$ tends to be close to zero, since the squared term in the denominator grows faster than the numerator. 


\subsection{Spatio-temporal random effect}\label{ST}
We decompose the latent random field $\xi_{ijt}$ into three components,
\begin{equation}
    \xi_{ijt} = \nu_t + \psi_t + \phi_{ijt}
    \label{eq:xi}
\end{equation}
In (\ref{eq:xi}),
$\nu_t$ is a sequence of independent and identically distributed Gaussian random effects with mean zero and variance $\sigma^2_{\nu}$. The term $\psi_t$ is a second-order random walk in time, capturing smooth long-term temporal trends. Finally, $\phi_{ijt}$ is a spatio-temporal random effect that evolves in time as a first-order autoregressive process:
\[\phi_{ijt} = \gamma\phi_{ij,t-1} + \omega_{ijt}; \qquad |\gamma| < 1\]
with Gaussian innovations $\omega_{ijt}$ that are independent across time and whose spatial dependence at each time point follows a Matérn covariance function with marginal variance $\sigma^2_{\omega}$, effective range $\rho$, and mean-square differentiability parameter 1.
 The spatio-temporal correlation function of $\xi_{ijt}$ implied by this decomposition is thus a combination of a purely temporal structured component $\psi_t$, an unstructured temporal component $\nu_t$, and a separable spatio-temporal component $\phi_{ijt}$
with Matérn spatial dependence and AR(1) temporal dependence. 

We did not include a purely spatial random effect, since $\phi_{ijt}$ already induces spatial dependence at each time point. Introducing a separate spatial-only term would add complexity with limited gains in predictive performance, based on preliminary exploratory analysis.

\section{Inference}\label{sec:inference}
\subsection{Predictor expression}
The sampling model is specified on the log scale, such that $\log S_{it}$ has a Gaussian distribution with mean $\mathbb{E}[\log S_{it}] = \mu_{S_{it}}$ and variance $\sigma^2_{S_{it}}$. The expression for $\mu_{S_{it}}$ is provided in Equation \eqref{eq:logsumexp}. Suppose that $\eta_{ijt} = \mu_{ijt}+\dfrac{1}{2}\sigma^2_Z$, then we can write $\mu_{S_{it}}$ as
\[
 \mathbb{E}[\log S_{it}] =\mu_{S_{it}} = \log \left( \sum_{j=1}^{J_i} \exp\big\{\eta_{ijt}\big\} \right) - \tfrac{1}{2}\sigma^2_{S_{it}} = \log \mathbb{E}[S_{it}] - \tfrac{1}{2}\sigma^2_{S_{it}}
\]
Thus, the predictor follows a \texttt{logsumexp} specification, with $-\tfrac{1}{2}\sigma^2_{S_{it}}$ acting as a moment-matching offset. The quantity $\tfrac{1}{2}\sigma^2_{S_{it}}$ is the difference between $\log \mathbb{E}[S_{it}]$ and $\mathbb{E}[\log S_{it}]$.


\subsection{Bayesian hierarchical specification}

Assuming conditional independence of the $S_{it}$ given all latent parameters, the full hierarchical specification of the model is:
\begin{equation}\label{eq:BHM}
    \begin{aligned}
        &S_{it}|\mu_{ijt},\sigma^2_Z,\bm{\theta} \overset{\text{ind}}{\sim} \log \text{Normal}\big( \mu_{S_{it}}, \sigma_{S_{it}}^2\big) \\
        &\mu_{S_{it}} = \log \sum_{j=1}^{J_i} \exp\left\{ \beta_0^* + \bm{\beta}^\intercal\bm{x}_{ijt} + \xi_{ijt}  \right\} - \tfrac{1}{2} \sigma^2_{S_{it}} \\
        &\begin{pmatrix} \beta_0^* & \bm{\beta} & \bm{\xi }\end{pmatrix}^\intercal|\bm{\theta} \sim \text{Normal}\big(\bm{0}, \inv{\mathbf{Q}(\bm{\theta})} \big)  \\
        &\bm{\theta}~\sim \pi(\bm{\theta})
    \end{aligned}
\end{equation}

where $\beta_0^* = \beta_0 + \tfrac{1}{2}\sigma^2_Z$, $\mathbf{Q}(\bm{\theta})$ is the precision matrix of the joint Gaussian distribution of the latent parameters $\begin{pmatrix} \beta_0^* & \bm{\beta} & \bm{\xi }\end{pmatrix}^\intercal$, and $\bm{\theta}$ collects all the latent field hyperparameters, with $\pi(\bm{\theta})$ as its prior distribution.

 Equation \eqref{eq:BHM} shows that $\beta_0$ and $\sigma^2_{Z}$ enter only through their sum, and are therefore not individually identifiable from the aggregated observations alone. We treat $\beta_0^*$ as a single parameter throughout. This does not affect the identifiability of the covariate effects 
$\bm{\beta}$ or the spatio-temporal random effects. Also, predictions at arbitrary spatial configurations remain valid under this reparametrisation, as discussed in Section \ref{subsec:pred_new_spatialconfigs}.

\subsection{Iterative estimation}\label{subsec:iterative}

Since $\sigma^2_{S_{it}}$ depends on both $\mu_{ijt}$ and $\sigma^2_Z$ through the moment-matching expressions, it cannot be computed analytically prior to model-fitting. Moreover, $\sigma^2_Z$ is not identifiable. Thus, we adopt the following iterative procedure:

\begin{enumerate}[label=Step~\arabic*:, wide=0pt, leftmargin=*]
    \item Fit the model ignoring $\dfrac{1}{2}\sigma^2_{S_{it}}$, i.e., setting $\sigma^2_{S_{it}}=0$ in the predictor expression for $\mu_{S_{it}}$ and assuming initially that the variance of the likelihood is constant for all $C_i$ and $t$.
    \item Consider a reasonable set of values for $\sigma^2_Z$, and evaluate the corresponding moment estimators for $\sigma^2_{S_{it}}$ (Equation \eqref{eq:logsumexp}).  We can write $\sigma^2_{S_{it}}$ as
    \[
    \sigma^2_{S_{it}}=\log \Bigg( \dfrac{\sum_{j=1}^{J_i} \Big(\exp\big\{\sigma^2_Z\big\}-1\Big)\Big(\exp\big\{ 2\eta_{ijt}\big\}\Big)  +  2\sum_{j<k}\exp\big\{ \eta_{ijt} + \eta_{ikt} \big\}\times\Big[\exp\big\{c_{ijk,t}\big\}-1\Big]}{\Big( \sum_{j=1}^{J_i}\exp\Big\{\eta_{ijt}\Big\} \Big)^2} + 1\Bigg)
    \]
    We evaluate $\sigma^2_{S_{it}}$ at the posterior means of $\eta_{ijt},\eta_{ikt},$ and $c_{ijk,t}$. In particular, suppose that $\begin{pmatrix} \hat{\beta}_0^* & \hat{\bm{\beta}} & \hat{\bm{\xi }}\end{pmatrix}^\intercal$ are the posterior means of the latent parameters, and that $\mathbf{A}$  denotes the known design matrix whose columns contain the covariates and the basis-function evaluations used to construct the Gaussian field $\bm{\xi}$ (see Section \ref{subsec:linearisedINLAandSPDE}). Then, the vector of estimated values of $\eta_{ijt}$'s is $\mathbf{A} \begin{pmatrix} \hat{\beta}_0^* \\ \hat{\bm{\beta}} \\ \hat{\bm{\xi }}\end{pmatrix}$. Similarly, the corresponding variances and covariances $c_{ijt,k}$ are obtained from $\mathbf{A} \inv{\mathbf{Q}(\hat{\bm{\theta}})} \mathbf{A}^\intercal$, where $\mathbf{Q}(\hat{\bm{\theta}})$ is the latent precision matrix evaluated at the posterior mode of $\bm{\theta}$. Because this matrix operation can be computationally expensive for large $\mathbf{Q}(\bm{\theta})$, we exploit the fact that only the covariance structure within each catchment $C_i$ and time $t$ is required, and therefore carry out the computations separately for each $(i,t)$ combination. Since $\sigma^2_{S_{it}}$ is evaluated using $\mathbf{Q}(\hat{\bm{\theta}})$ together with  posterior point estimates of the latent parameters, this corresponds to an empirical Bayes approximation.
    
    \item Among the candidate $\sigma^2_{S_{it}}$ estimates, consider the maximum value (for each catchment $C_i$ and time $t$) across the $\sigma^2_Z$ values considered in Step 2. This is a reasonable choice since it caters for maximum uncertainty in the latent model. 
    \item Refit the model incorporating the estimated values of $\sigma^2_{S_{it}}$ in Step 3 as an offset in the predictor expression for $\mu_{S_{it}}$ and as a fixed scaling parameter for the likelihood precision.
    \item Repeat Steps 2--4 until convergence, i.e. until the change in each of the posterior estimates of latent parameters between successive iterations  falls below a chosen tolerance threshold.
\end{enumerate}


\subsection{Linearised INLA and SPDE approximation}\label{subsec:linearisedINLAandSPDE}

We use the Integrated Nested Laplace Approximation (INLA) \citep{rue2009approximate, van2023new} to obtain the posterior distribution of all unknown quantities,
\[
\pi(\beta_0^*,\bm{\beta},\bm{\xi},\bm{\theta},\sigma^2_{S_{it}}|\textbf{S}).
\]
Since the predictor expression for $\mu_{S_{it}}$ in Equation \eqref{eq:BHM} is a nonlinear function of the latent parameters $\beta_0^*,\bm{\beta}$, and $\bm{\xi}$, the standard INLA
implementation cannot be applied directly. We therefore use  
the linearised INLA approach \citep{lindgren2024inlabru,suen2026coherent}, which handles nonlinear predictor expressions through an approximate iterative method based on a first-order Taylor expansion of the nonlinear predictor, applying INLA at each linearised model configuration. We implement model estimation using the \texttt{inlabru} package. 

To handle the spatially continuous latent Gaussian field $\bm{\xi}$, we use the Stochastic Partial Differential Equations (SPDE) approach \citep{lindgren2011explicit}, which approximates the Matérn Gaussian field by a Gaussian Markov random field defined on a triangulated mesh, enabling computationally efficient inference. The mesh used for the SPDE approximation is shown in Figure \ref{fig:mesh}. Prior specifications for all  model parameters are presented in Section \ref{sec:application}.

\section{Application to wastewater virus concentrations}\label{sec:application}

\subsection{Outcome data}

The outcome data consist of weekly measurements of SARS-CoV-2 N1 virus concentration (gene copies per liter) at 47 catchment areas in Wales (see Figure \ref{fig:lsoa_catchment_intersection}) covering approximately 80\% of the welsh population \cite{hoffmann2023wastewater}. The study period spans the weeks commencing 2022-08-02 to 2023-07-25, providing $T=52$ weekly measurements for each catchment area. A heat map of the data is shown in Figure \ref{fig:heatmap} in the Appendix. Following \citep{Roberts2022RTqPCR}, we applied flow normalization to the virus concentrations. Specifically, each observed concentration (gene copies per unit volume) was multiplied by the corresponding wastewater flow rate to estimate the total number of gene copies passing through the sampling point over the measurement period in question. This transformation accounts for variations in wastewater flow, such as those caused by rainfall, allowing the outcome variable
to better represent the total viral load. The data contain missing values  for some catchment-time combinations, as can be seen in Figure \ref{fig:heatmap}. These are treated as missing completely at random and handled automatically within the Bayesian inferential framework by  conditioning only on observed data.

\subsection{Covariate construction}

\begin{figure}
    \centering
    \includegraphics[width=0.8\linewidth,
        trim=0cm 0cm 5cm 0cm, clip]{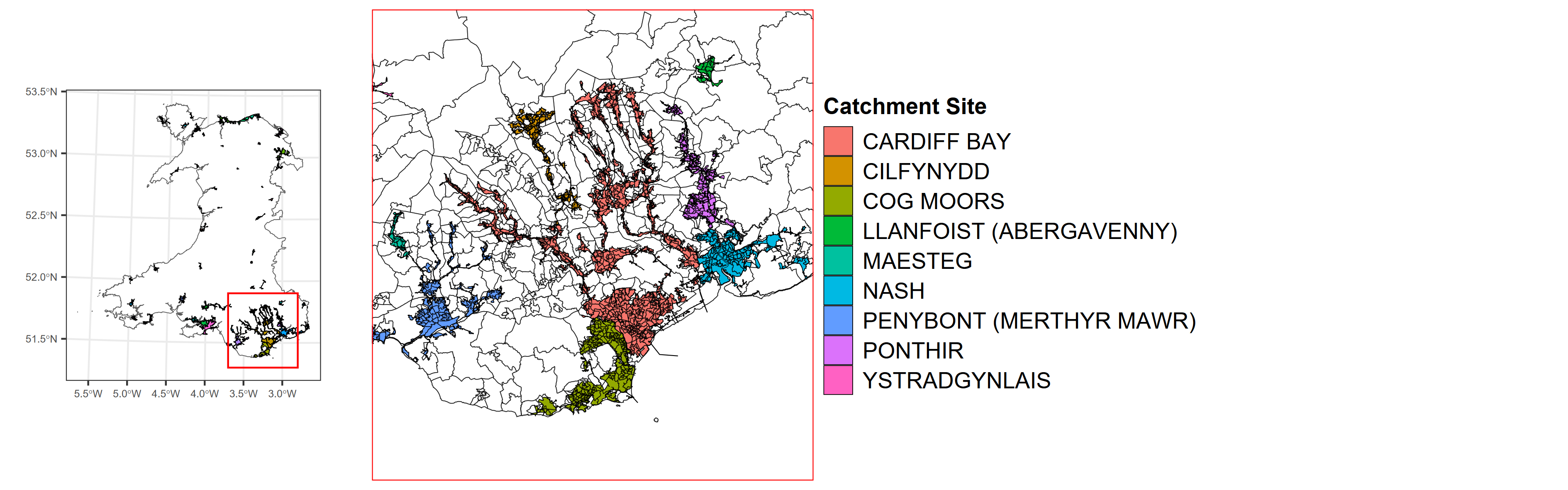}
    \caption{LSOA-catchment intersection. Shown in the left plot are the catchment areas for Wales, while in the right is a zoomed in portion of Cardiff Area. The largest catchment is shown in the right plot, which is Cardiff Bay Swk.} 
    \label{fig:lsoa_catchment_intersection}
\end{figure}

\begin{figure}
    \centering
    \includegraphics[width=0.95\linewidth]{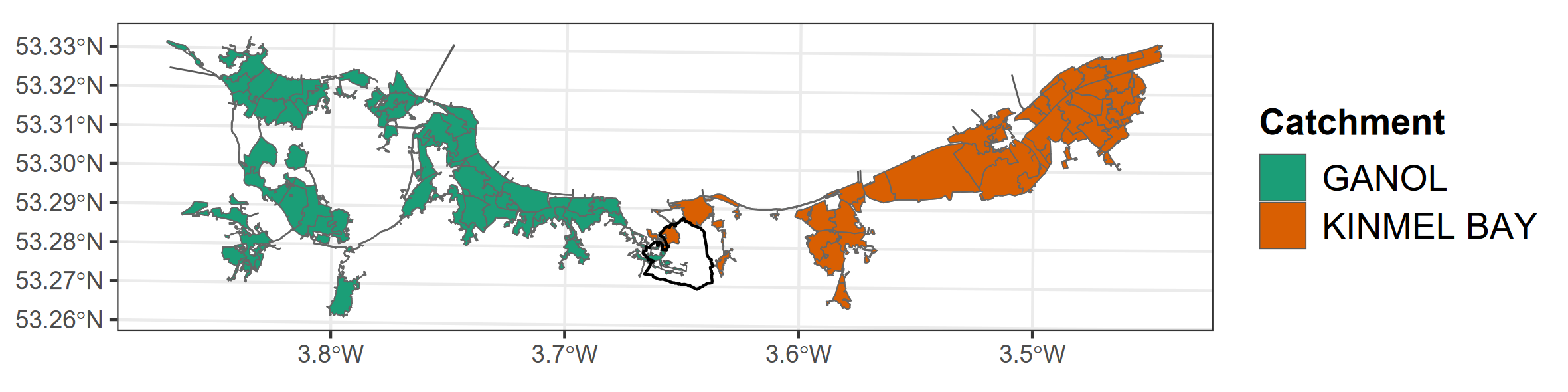}
    \caption{An LSOA (in solid black line) serviced by two catchment sites: Ganol and Kinmel Bay.}
    \label{fig:lsoa_catchment_intersection_2}
\end{figure}

We assume that $S_{it}$ is an aggregated value from the nested intersections between Lower Super Output Areas (LSOAs) $\ell_j, j=1,\ldots, 1917$ and the catchment areas $C_i, i=1,\ldots, 47$. Figure \ref{fig:lsoa_catchment_intersection} shows these area intersections between the $C_i$ (shown with colours) and the $\ell_j$ (shown with solid black boundaries). Figure \ref{fig:lsoas_per_catchment} in the Appendix shows the number of LSOAs served by each catchment site. Cardiff Bay is the largest catchment site, serving almost 400 LSOAs, whereas more than half of the catchment sites serve fewer than 20 LSOAs. Out of 1,917 LSOAS, only 1,407 intersect at least one catchment area, while 60 LSOAs intersect two catchments. For example, Figure \ref{fig:lsoa_catchment_intersection_2} shows an LSOA (in black solid line) serviced by two catchment sites: Ganol and Kinmel Bay. These intersections remain within the block aggregation framework, as each LSOA–catchment intersection defines a well-defined spatial block over which the underlying continuous process is aggregated.  Consequently, it is important to derive geospatial estimates of the covariates included in the analysis for the LSOA-cathment intersections. The covariates considered are population count and land cover, and the methods used to obtain these estimates are described below.

\paragraph*{\textbf{Population counts}.} 
Following \cite{hoffmann2023wastewater}, let $b_{ij}$ be the overlap (spatial intersection) between catchment $C_i$ and LSOA $\ell_j$. We allocate the population of each LSOA to its intersecting catchments according to the relative area of these intersection blocks. Specifically, the population associated with the intersection $b_{ij}$ at time $t$ is estimated as
 \begin{equation}\label{eq:popestimates}
     Pop_{ijt}
     =
     \frac{|b_{ij}|}{\sum_{i'} |b_{i'j}|}
     Pop(\ell_j,t),
 \end{equation}
 
where $|b_{ij}|$ denotes the surface area of the LSOA–catchment intersection, and $Pop(\ell_j,t)$ denotes the population of LSOA $\ell_j$ at time $t$ obtained from annual census estimates. The denominator $\sum_{i'}|b_{i'j}|$ represents the total area of LSOA $\ell_j$ that overlaps the catchment areas. Thus, $Pop_{ijt}$ provides an area-weighted estimate of the population associated with each LSOA–catchment intersection and ensures that the LSOA population is allocated across its intersecting catchments according to their relative spatial overlap. For LSOAs intersecting a single catchment, this allocation assigns the entire LSOA population to that catchment, whereas for LSOAs intersecting multiple catchments, the population is partitioned according to the relative areas of the intersection blocks.


\paragraph*{\textbf{Land cover}.} 
We also include the proportion of each LSOA-catchment intersection $b_{ij}$ classified as urban ($\% Urban$) as a covariate, computed from land cover data. Figure 3b shows the $\% Urban$ covariate by LSOA.

\subsection{Model specification and priors}

Following the general framework in Section \ref{sec:model}, the predictor expression for the latent mean at LSOA-catchment intersection $b_{ij}$ and time $t$ is:
\[\mu_{ijt}=\beta_0 + \beta_1 Pop_{ijt} + \beta_2 \%Urban_{ij} + \xi_{ijt}\]
where $Pop_{ijt}$ is the geospatial population estimate defined in Section  4.2, $\%Urban_{ij}$ is the proportion of $b_{ij}$ classified as urban, and $\xi_{ijt}$ is the spatio-temporal random effect specified in Section \ref{ST}.

\begin{figure}[]
    \centering

    \begin{subfigure}{0.42\linewidth}
        \centering
        \includegraphics[width=\linewidth]{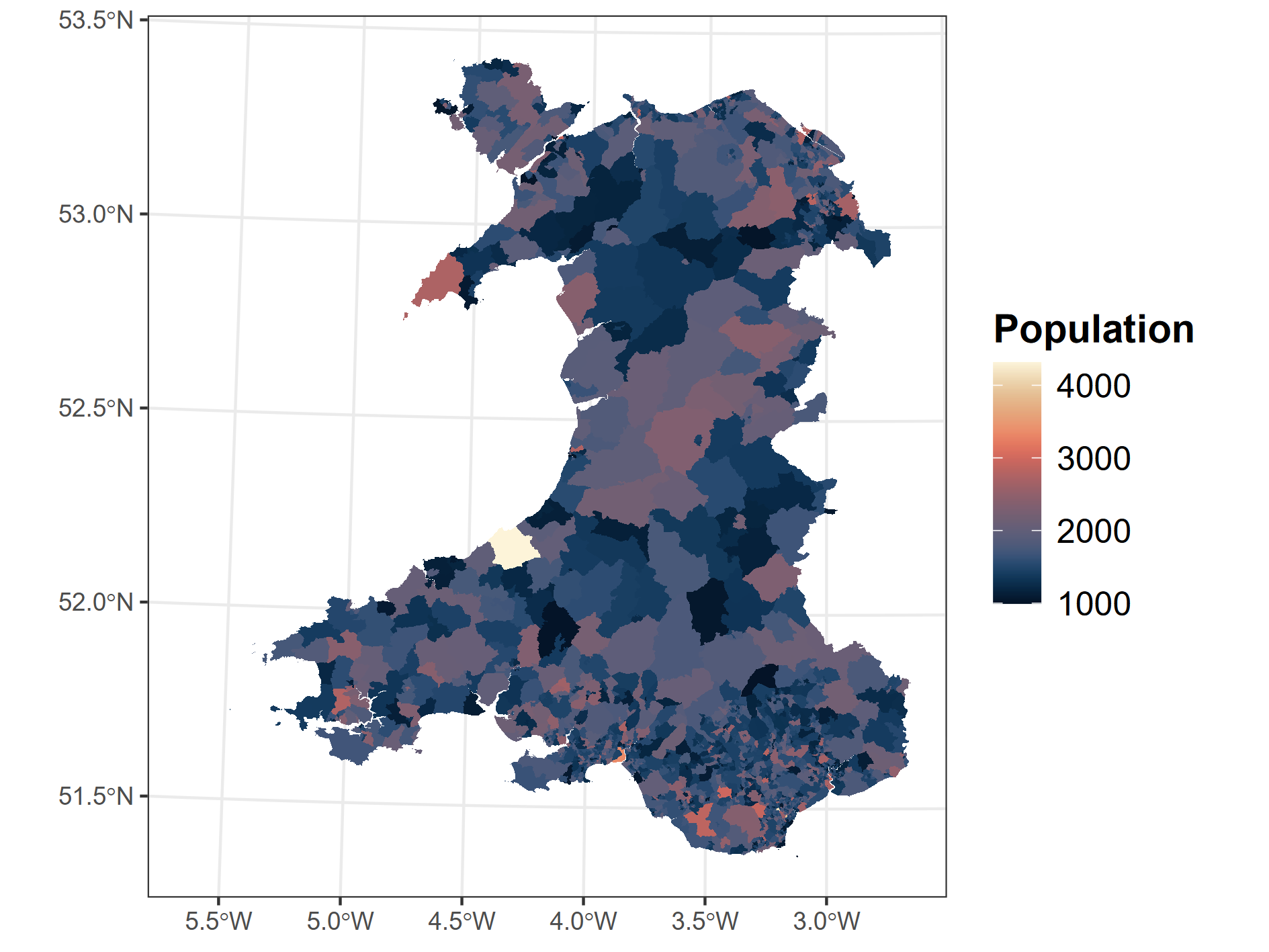}
        \caption{Population by LSOA}
    \end{subfigure}
    \begin{subfigure}{0.42\linewidth}
        \centering
        \includegraphics[width=\linewidth]{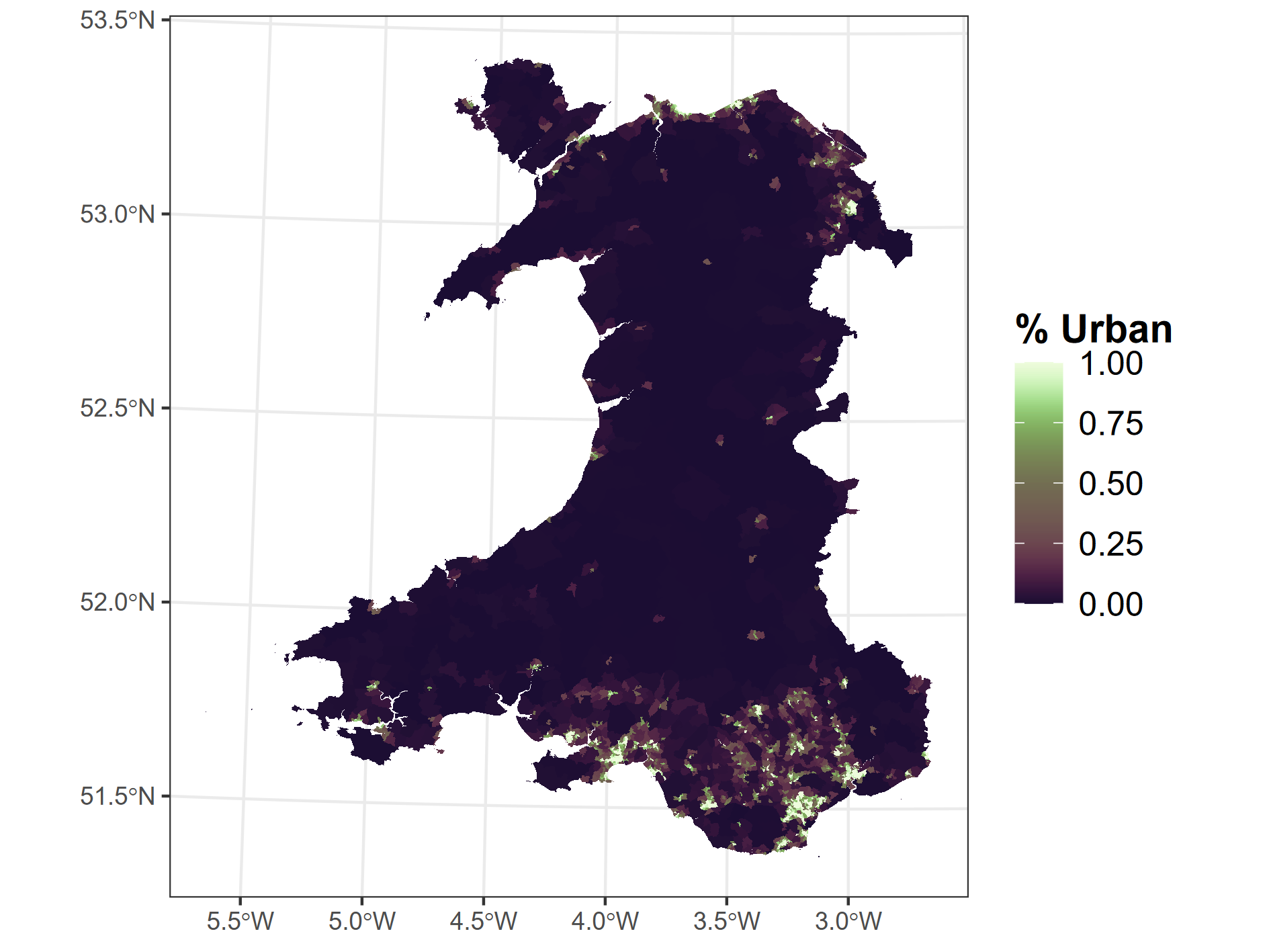}
        \caption{\% Urban by LSOA}
    \end{subfigure}

    \caption{Covariates: (a) population of LSOAs in Wales, (b) \% of land cover considered urban for each LSOA}
\end{figure}
\begin{figure}[]
    \centering
 \begin{subfigure}{0.37\linewidth}
        \centering
        \includegraphics[width=\linewidth]{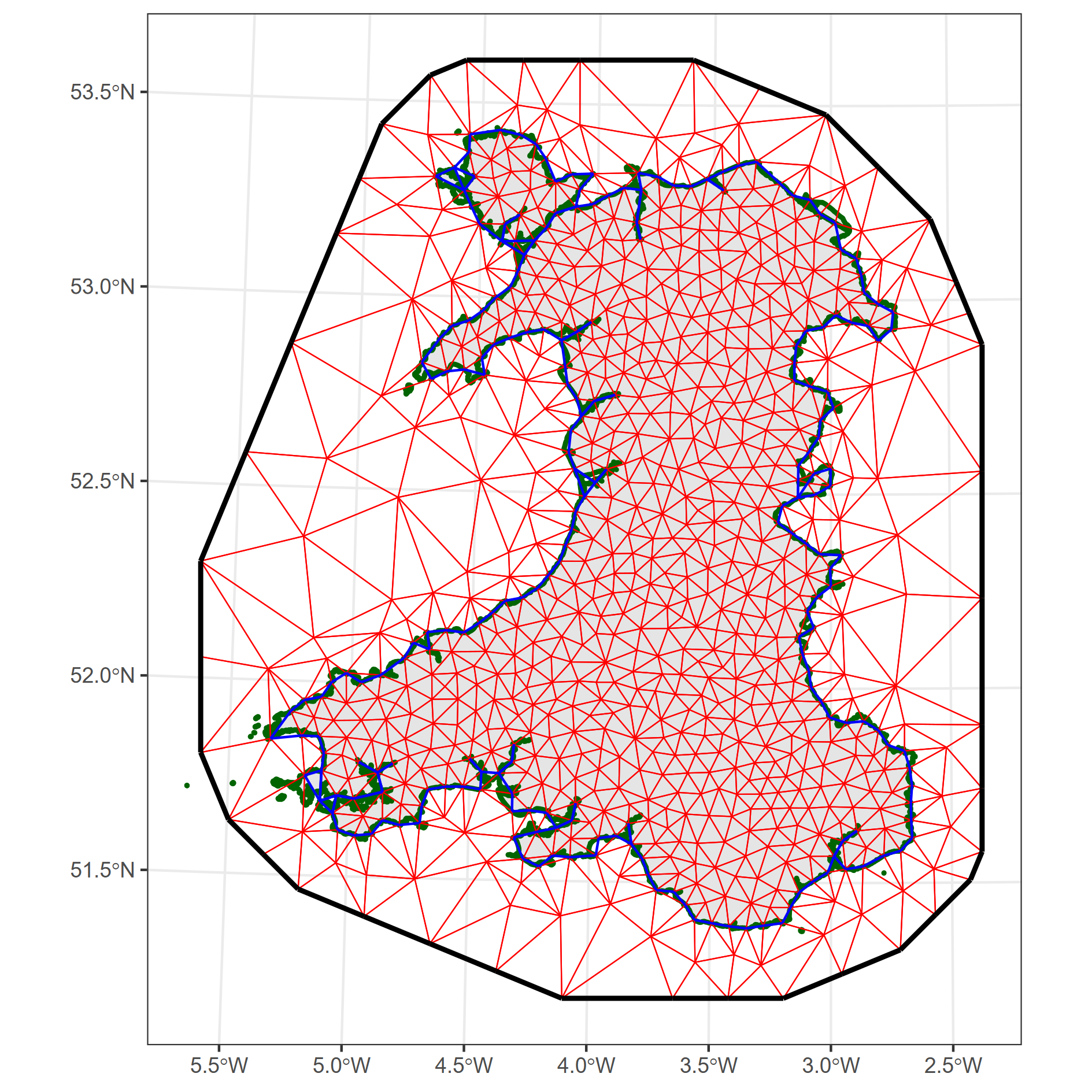}
        \caption{Mesh for SPDE model}
        \label{fig:mesh}
    \end{subfigure}
    \begin{subfigure}{0.52\linewidth}
        \centering
        \includegraphics[trim={2cm 0 2cm 0}, clip,width=\linewidth]{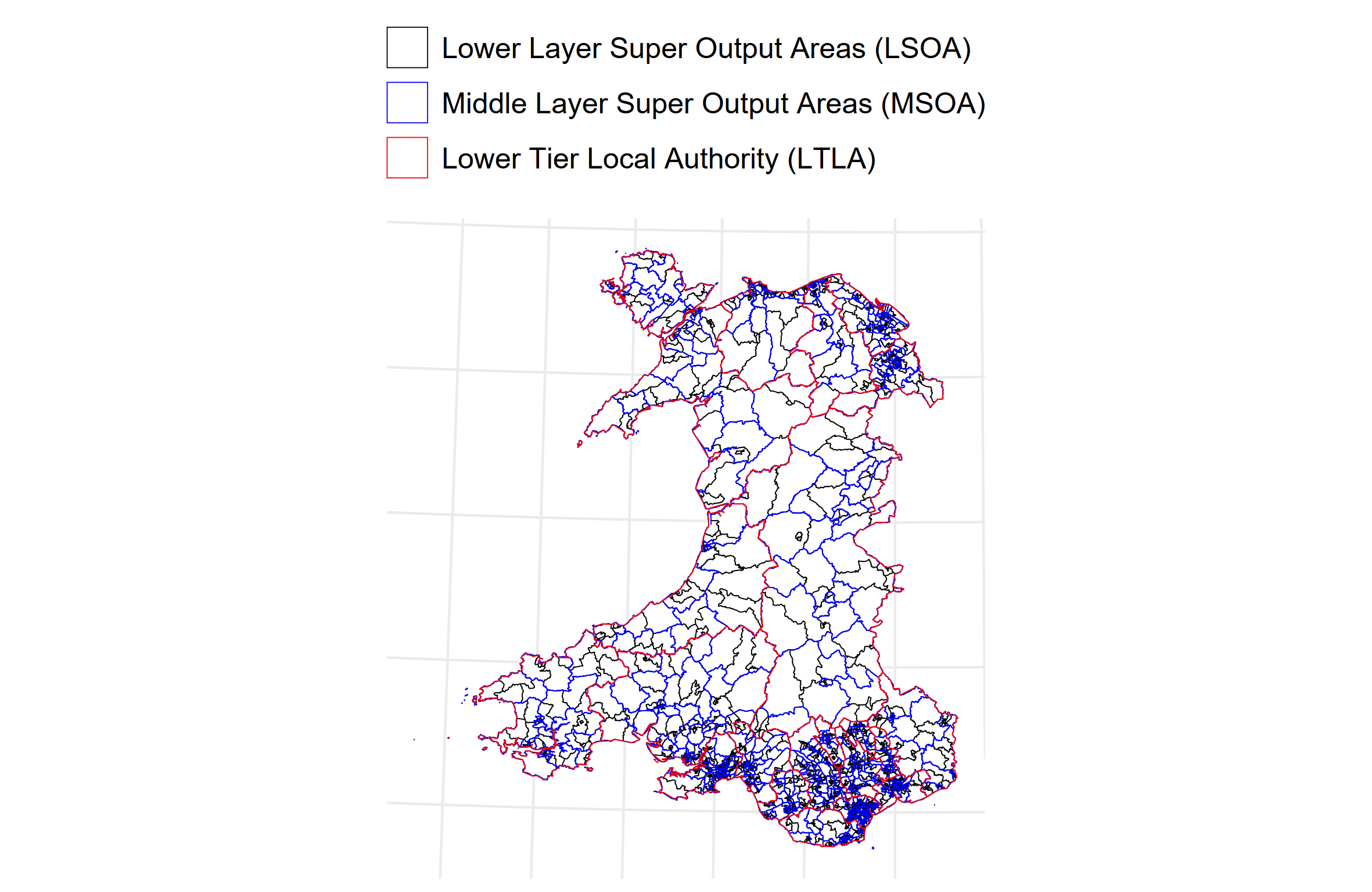}
        \caption{Different areal boundaries in Wales}
        \label{fig:wales_lsoa_msoa_ltla}
    \end{subfigure}

    \caption{(a) mesh used for the SPDE model, (b) spatial boundaries of LSOAs, MSOAs, and LTLAs in Wales}
\end{figure}

For the initial model, in which the $\sigma^2_{S_{it}}$ terms are ignored, we specify diffuse priors for $\beta_0^*$ and $\bm{\beta}$: $\beta_0^* \sim \text{Normal}(0,\infty)$ and $\bm{\beta} \sim \text{Normal}(\bm{0},\infty)$. For the subsequent model estimation, we specify an informative Gaussian prior for $\beta_0^*$, with its mean set to the posterior mean estimate from the initial model and its variance set to 2. For the AR(1) parameter $\gamma$, we have $\log\Big( \tfrac{1+\gamma}{1-\gamma}\Big)\sim\text{Normal}(0,0.15^2)$. We specify penalized-complexity (PC) priors for the parameters of the Gaussian field $\xi_{ijt}$. PC priors provide a principled Bayesian framework for controlling model complexity by shrinking the model towards a clearly defined and interpretable base model, such as zero standard deviation for random effects or infinite spatial range for spatial processes \citep{simpson2017penalising, fuglstad2019constructing}. The priors are calibrated through interpretable probability statements, allowing weakly informative prior beliefs about plausible effect sizes to be incorporated. For the initial model, we specify the following PC priors: $\mathbb{P}(\sigma_{\omega}>1.97)=0.5$ and $\mathbb{P}(\rho<86 \text{km})=0.5$. The prior for $\rho$ reflects the expectation that the effective spatial range of the viral concentration process is on the order of tens of kilometres, consistent with the spatial scale of the catchment areas in Wales. The prior for $\sigma_{\omega}$ reflects a weakly informative belief about the magnitude of spatial variation in viral load. For the subsequent parameter estimation, the posterior mean estimates of $\sigma_{\omega}$ and $\rho$ from the initial model are used to calibrate the corresponding PC priors, i.e., $\mathbb{P}(\sigma_{\omega}>\hat{\sigma}_{\omega,post})=0.5$ and $\mathbb{P}(\rho<\hat{\rho}_{post})=0.5$, where $\hat{\sigma}_{\omega,post}$ and $\hat{\rho}_{post}$ are the posterior mean estimates from the initial model. Finally, for the variance parameters of the temporal random effects $\nu_t$ and $\psi_t$, we specify inverse-Gamma$(1,0.00005)$ priors for the initial model. Following estimation of the initial model, we use the posterior mean estimates of the corresponding standard deviations to calibrate PC priors for the subsequent model parameter estimation. Specifically, the PC priors are defined by
$\mathbb{P}(\sigma_{\nu} > \hat{\sigma}_{\nu,\mathrm{post}})=0.5,
\mathbb{P}(\sigma_{\psi} > \hat{\sigma}_{\psi,\mathrm{post}})=0.5$,
where $\hat{\sigma}_{\nu,\mathrm{post}}$ and $\hat{\sigma}_{\psi,\mathrm{post}}$ denote the posterior mean estimates of the standard deviations of $\nu_t$ and $\psi_t$, respectively, obtained from the initial model.




\subsection{Predictions at new administrative boundaries}\label{subsec:pred_new_spatialconfigs}

For downstream analysis, for example using wastewater model predictions as metrics of disease incidence, a prerequisite is to compute wastewater virus metrics at new spatial configurations, such as Middle Layer Super Output Areas (MSOA) or Lower Tier Local Authority (LTLA). Figure \ref{fig:wales_lsoa_msoa_ltla} shows the boundaries of the three administrative partitions: LSOAs, MSOAs, and LTLAs. 

The proposed model supports predictions at arbitrary spatial configurations. Since LSOAs $\ell_j$ are nested within the MSOAs, which in turn are nested within LTLAS, the aggregation is straightforward.  

Let $Z(\ell_j,t)$ denote the unobserved total number of gene copies in LSOA $\ell_j$ and week $t$. For LSOAs that are serviced by at least one STW, the predicted value is
a sum of the individual contributions from the nested subareas serviced by different STWs,
\begin{equation}\label{eq:pred_LSOA}
    \begin{aligned}
        \mathbb{E}\big[Z(\ell_j,t)\big] &= \sum_{b_{ij}: b_{ij}\cap \ell_j\neq\phi} \mathbb{E}\big[Z_{ijt}\big] \\
        & =\sum_{b_{ij}: b_{ij}\cap \ell_j\neq\phi} \exp\big\{\beta_0^* + \bm{\beta}^\intercal\bm{x}_{ijt} + \xi_{ijt}\big\}.
    \end{aligned}
\end{equation}
 For LSOAs not serviced by any STW, predictions are obtained directly from the latent model as
 $$\mathbb{E}\big[Z(\ell_j,t)\big] = \exp\big\{\beta_0^* + \bm{\beta}^\intercal\bm{x}_{\ell_j t} + \xi_{\ell_j t}\big\},$$ where $\bm{x}_{\ell_j,t}$ is the vector of the LSOA-level covariates and $\xi_{\ell_j t}$ is the predicted latent field at the LSOA centroid. 

Predictions at the MSOA and LTLA levels are obtained by aggregation:
\begin{equation}\label{eq:pred_MSOA_LTLA}
   \mathbb{E}\big[Z(\text{MSOA}_x,t)\big] =  \sum_{b_{ij}: b_{ij}\cap \text{MSOA}_x\neq\phi }\mathbb{E}\big[Z(\ell_j,t)\big], \;\;\;\;\; \mathbb{E}\big[(\text{LTLA}_y,t)\big] = \sum_{b_{ij}: b_{ij}\cap \text{LTLA}_y\neq\phi}\mathbb{E}\big[Z(\ell_j,t)\big].
\end{equation}

The uncertainty of the predictions at new administrative boundaries is obtained via posterior sampling.  To sample from the predictive distributions at the new block configurations, we first sample from the posterior distributions of all unobserved quantities, and then evaluate $\mathbb{E}\big[Z(\ell_j,t)\big]$, $\mathbb{E}\big[Z(\text{MSOA}_x,t)\big]$ and $\mathbb{E}\big[(\text{LTLA}_y,t)\big]$ using each sample. 


\section{Results}\label{sec:results}

All figures in this section are presented on the base-10 logarithmic scale  unless otherwise stated. Model estimation was performed using natural logarithm (base e), consistent with the \texttt{logsumexp} model specification, and results have been rescaled accordingly for presentation. Section \ref{subsec:resultsiterative} presents summary results from the iterative estimation approach, focusing on convergence. Section \ref{subsec:results_estimates} presents the estimated model parameters. We performed cross-validation to assess the capability of the proposed model for spatial prediction. These are  presented in Section \ref{subsec:lgocv}. Section \ref{subsec:pred_new_spatialconfigs} presents predictions at new administrative boundaries. Section \ref{subsec:link_ww_pcr} presents the results of an exploratory analysis linking wastewater viral load at the LTLA level with positivity rates from PCR tests.

\subsection{Iterative estimation}\label{subsec:resultsiterative}

\begin{figure}
    \centering
    \includegraphics[width=.75\linewidth]{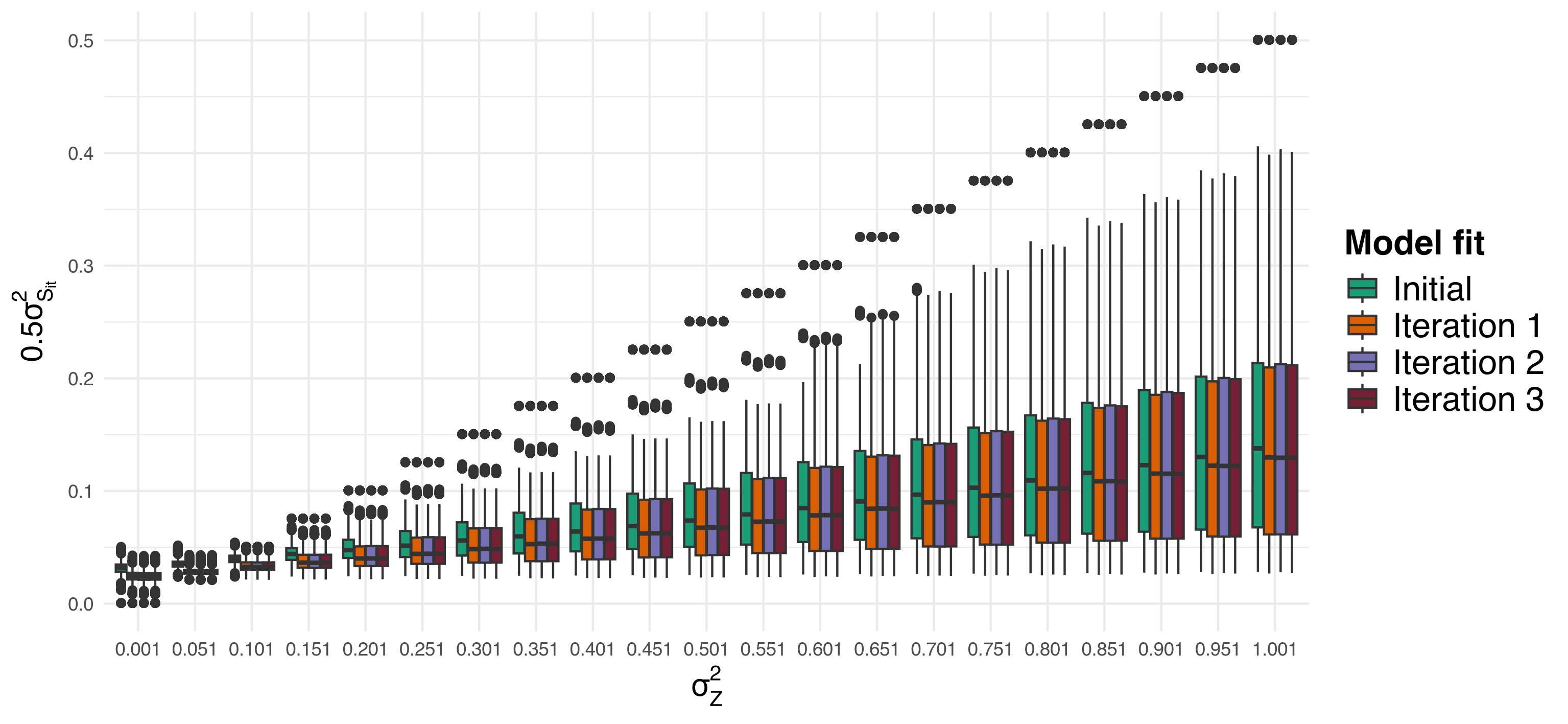}
    \captionof{figure}{Distribution of the estimated values of $\dfrac{1}{2}\sigma^2_{s_{it}}$ for different values of $\sigma^2_Z$ across iterations.}
    \label{fig:sensitivity_plot_sigma}
\end{figure}

For the iterative estimation described in Section \ref{subsec:iterative}, we considered $\sigma^2_Z$ values over
the range 0.001 to 1. These are the empirical values of the within-catchment variation in the latent $\mu_{ijt}$ values. Figure \ref{fig:sensitivity_plot_sigma} shows the distribution of the estimated values of $\tfrac{1}{2}\sigma^2_{S_{it}}$ with respect to $\sigma^2_Z$ up to the third iteration. The estimates appear to be stable even after the first iteration. The maximum change in the estimated $\sigma^2_{S_{it}}$ is 1.89\% after the third iteration. Moreover,  $\frac{1}{2}\sigma^2_{S_{it}}$ remains small even at the maximum value of $\sigma^2_Z$ considered. The estimated values of $\frac{1}{2}\sigma^2_{S_{it}}$ are negligible relative to the scale of the log-transformed (natural logarithm) viral load, which ranges from 19.18 to 31.59 units. Consequently, incorporating these terms as an offset in the predictor expression for $\mu_{S_{it}}$ does not materially affect the posterior estimates.

\begin{figure}
    \centering
    \includegraphics[width=0.75\linewidth]{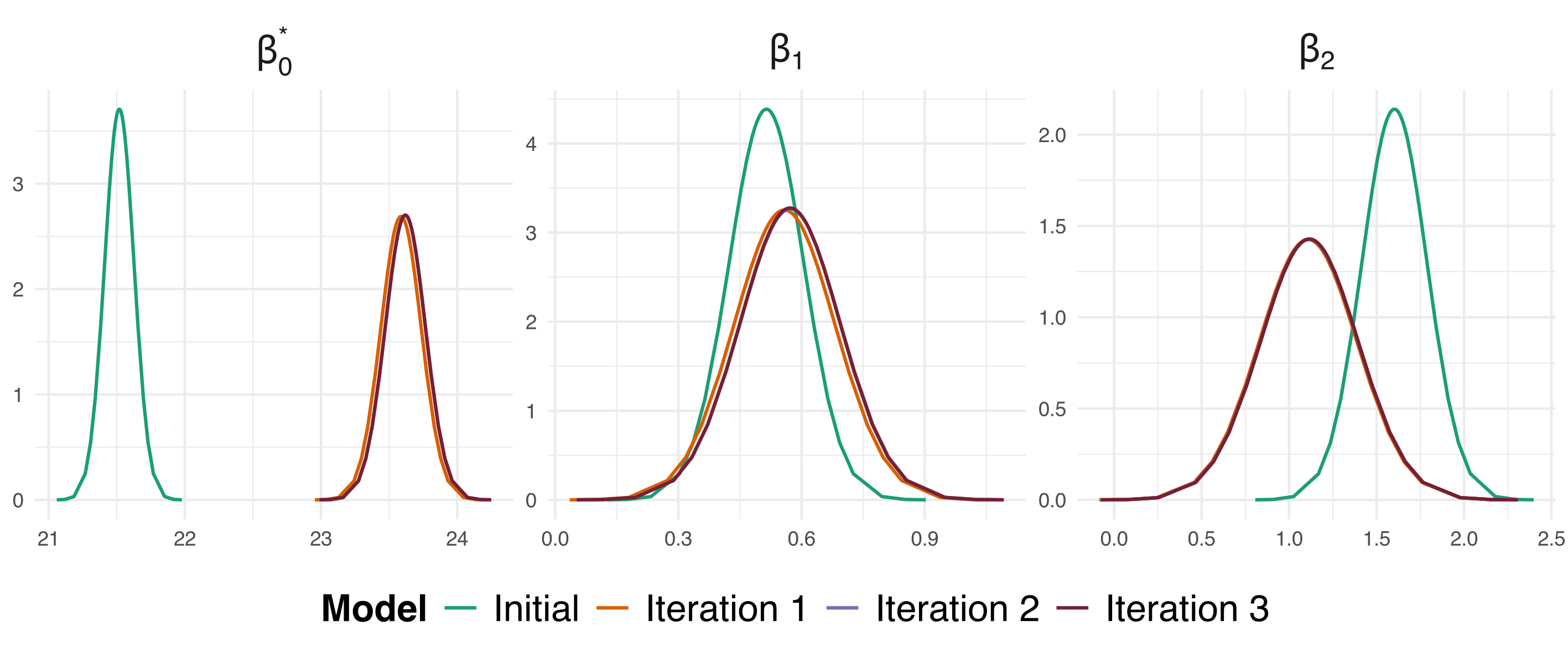}
    \captionof{figure}{Comparison of the posterior distributions of fixed effects for the first two iterations.}
    \label{fig:fixedeffectscompare}
\end{figure}


\subsection{Fixed effects and random effects estimates}\label{subsec:results_estimates}


Figure \ref{fig:fixedeffectscompare} shows the posterior distributions of the fixed effects $\beta_0^*,\beta_1$, and $\beta_2$. Only negligible changes are observed after the first iteration, suggesting that the fixed effects estimates have effectively stabilised. Table \ref{tab:modelestimates} presents posterior estimates of model parameters. Both covariates are positively associated with the log number of gene copies. The posterior mean and 95\% credible interval  for the population effect is 0.567 (0.328, 0.805) indicating that sub-areas with larger populations contribute more viral load to the catchment, as expected. The posterior mean and 95\% credible interval for the urban land cover effect is 1.103 (0.556, 1.651) suggesting that more urbanised sub-areas are associated with substantially higher gene copy counts, potentially reflecting higher population density, greater connectivity of the sewer network, or differences in viral shedding behaviour across urban and rural populations. The estimated effective spatial range of the Matérn random field is approximately 37 km (20.256 km, 57.769 km) which is on the order of the spatial scale of the catchment areas in Wales, suggesting that spatial dependence in viral load operates at a scale commensurate with the catchment infrastructure. The AR(1) parameter is estimated as 0.985 (0.971, 0.994) indicating strong temporal persistence in the spatio-temporal random effect, i.e. the viral load at a given location in a given week is strongly predictive of the following week's value. The random walk component $\psi_t$ has a higher marginal variance than the the iid component $\nu_t$, consistent with $\psi_t$ capturing smoother, longer-term temporal trends, while $\nu_t$ captures short-term fluctuations. 




\begin{table}
\centering
\scalebox{1}{\begin{tabular}{|l|rrrr|}
  \hline
  \textbf{Parameter} & \textbf{Mean} & \textbf{SD} & \textbf{P$2.5^{\text{th}}$} & \textbf{P$97.5^{\text{th}}$} \\ \hline

  $\beta_0^*$ & 23.603 & 0.148 & 23.314 & 23.893  \\
  $\beta_1$, Population & 0.567 & 0.122 & 0.328 & 0.805 \\
  $\beta_2$, \% Urban & 1.103 & 0.279 & 0.556 & 1.651 \\ \hline

  $\sigma^2_{\nu}$  & 0.064 & 0.022 & 0.031 & 0.115 \\
  $\sigma^2_{\psi}$ & 0.118 & 0.026 & 0.076 & 0.177 \\
  $\rho$ (km)       & 36.123 & 9.606 & 20.256 & 57.769 \\
  $\sigma_{\omega}$ & 0.554 & 0.062 & 0.441 & 0.686   \\
  $\phi$            & 0.985 & 0.006 & 0.971 & 0.994  \\ \hline

\end{tabular}}
\caption{Posterior estimates for the wastewater virus model after the 3rd iteration.}
\label{tab:modelestimates}
\end{table}

Figure \ref{fig:temporaleffect} displays the estimated residual temporal effects, $\nu_t + \psi_t$, plotted alongside the log total gene copies for Wales. The two curves show strikingly similar temporal patterns. There is a noticeable rise in temporal effects leading up to the Christmas holiday period, followed by a decline at the start of the new year. A further decrease is observed around the middle of 2023.  This close agreement between the pattern of variation in the estimated temporal effects and the observed aggregate signal provides informal validation of the model's ability to capture the dominant temporal dynamics of SARS-CoV-2 transmission in Wales.

\begin{center}
    \includegraphics[width=0.9\linewidth]{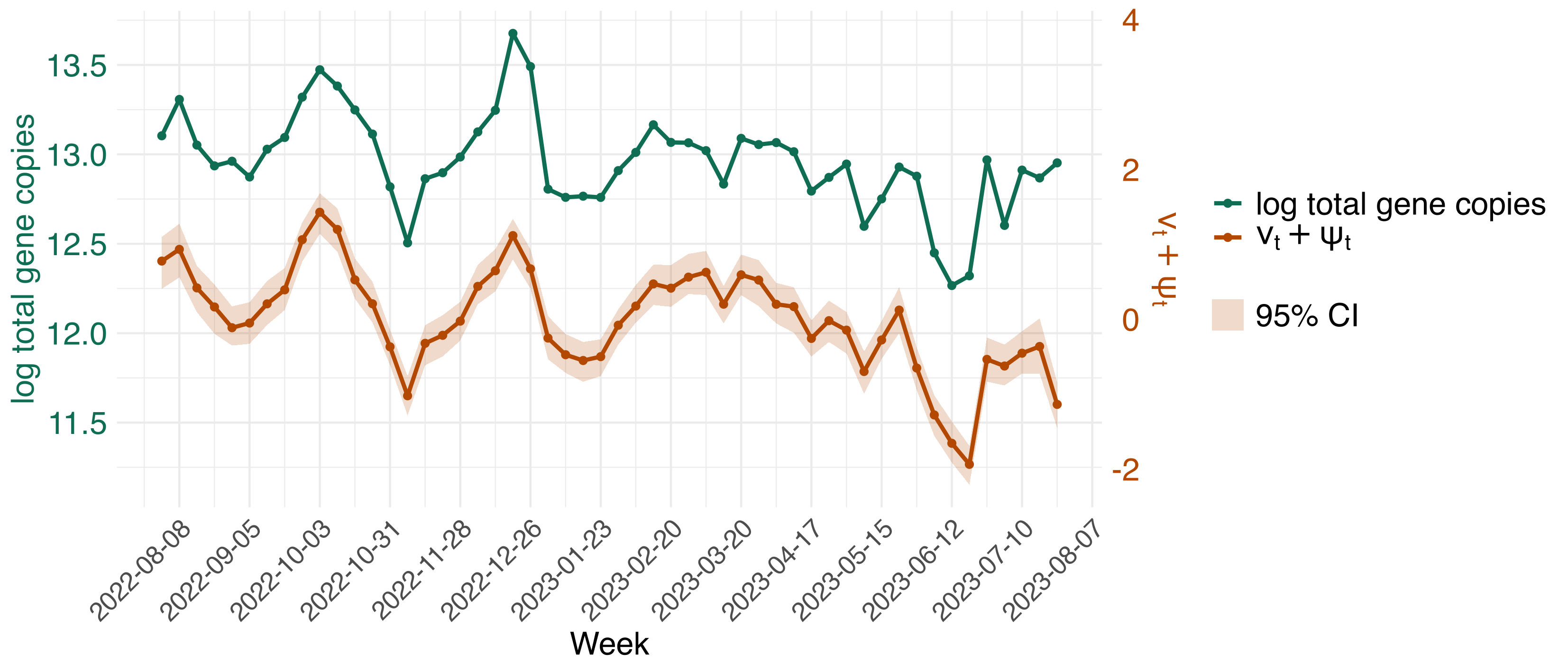}
    \captionof{figure}{Plot of observed log total gene copies for Wales versus the estimated residual temporal effect ($\nu_t+\psi_t$).}
    \label{fig:temporaleffect}
\end{center}

\subsection{Leave-group-out crossvalidation}\label{subsec:lgocv}

We assess the predictive performance of the model using leave-group-out cross-validation (LGOCV)  \citep{adin2024automatic}. When generating a prediction for $S_{it}$, we remove not only all data points from the catchment $C_i$, but also the data from the 4 nearest catchment areas. This approach is used to avoid artificially inflating predictive performance through spatial smoothing over nearby observations.  

Predicting the values of $S_{it}$ for a catchment $C_i$ is performed using the following steps:
\begin{enumerate}[label=\textbf{Step \arabic*:}]
    \item Generate $B$ posterior samples of the $b_{ij}$-level latent linear predictor: $\eta_{ijt}^{(b)} = \beta_0^{^{*(b)}} + \bm{\beta}^{(b)\intercal}\bm{x}_{ijt} + \xi_{ijt}^{(b)}, b=1,\ldots,B$; where $\beta_0^{^{*(b)}},\bm{\beta}^{(b)},$ and $\xi_{ijt}^{(b)}$ are samples from the estimated posteriors of the latent parameters. 
    \item Calculate the variance $\sigma^{2^{(b)}}_{S_{it}}$ for each $\eta_{ijt}^{(b)}$, using Equation \eqref{eq:first_second_moments_Sit}.
    \item Calculate $\mu_{S_{it}}^{(b)}$ for each $\eta_{ijt}^{(b)}$, given by: $\mu_{S_{it}}^{(b)} = \log \sum_{j=1}^{J_i} \exp\left\{ \eta_{ijt}^{(b)}  \right\} - \tfrac{1}{2} \sigma^{2^{(b)}}_{S_{it}}$.
    \item Sample from the predictive distribution: $S_{it}^{(b)}\sim \log\text{Normal}\big(\mu_{S_{it}}^{(b)},\sigma^{2^{(b)}}_{S_{it}}\big)$
    \item Combine all values $S_{it}^{(b)}, b=1,\ldots,B$. This is the estimated posterior predictive distribution of $S_{it}$. 
\end{enumerate}

The model achieves a coverage of 94.53\% and RMSE of 0.3950, whereas the values for the outcome variable range from 8.04 to 13.04 log (base 10) gene copies. Figure \ref{fig:lgocv_3sites_scenarioA} shows a comparison of the predicted and osbserved values for three catchment areas: Cardiff Bay Swk, which is the largest catchment area; Ganswllt, which shows the smallest prediction RMSE out of the 47 catchment areas; and Llanfyllin, which is the smallest catchment area. Predictions for Garnswllt align closely with the observed values, with observations well-contained within the 95\% prediction intervals. For Cardiff Bay Swk, predictions tend to be overestimated during the period from mid-March to May 2023. This period coincides with a localised decline in viral load that is not fully captured by the model, possibly reflecting site-specific factors or changes in population behaviour during this period. Full LGOCV prediction plots for all 47 catchment areas are provided in Figure \ref{fig:CVplotsall} in the Appendix.

Figure \ref{fig:combined_scatter_side_scenA} shows scatterplots of  RMSE and coverage (from LGOCV predictions) against the log catchment area size ($\text{m}^2$). The results indicate that larger catchments generally achieve lower RMSE values and higher coverage, suggesting better predictive performance. This is consistent with the expectation that larger catchments aggregate over more sub-areas, reducing the influence of local noise. Three small catchments show poor  coverage: Builth Wells (coverage 74.51\%, RMSE 0.71); Pwllheli (coverage 80.39\%, RMSE 0.64); and Cardigan (coverage 84.62\%, RMSE 0.59). These are also the three sites with high RMSEs. As shown in the LGOCV prediction plots in the Appendix (Figure \ref{fig:CVplotsall}), predictions for these sites tend to be overestimated, suggesting that the model may not adequately capture the dynamics of very small, potentially isolated catchments. Also notable in Figure \ref{fig:combined_scatter_side_scenA} is the largest site, Cardiff Bay Swk, which did not achieve a coverage as high as other large sites. Nevertheless, its RMSE remained low overall, based on Figure \ref{fig:lgocv_3sites_scenarioA}, although the predictions did not accurately capture the drop in observed viral load towards the beginning of the second quarter of 2023.

\begin{center}

\begin{minipage}{\linewidth}
    \centering
    \includegraphics[width=.9\linewidth]{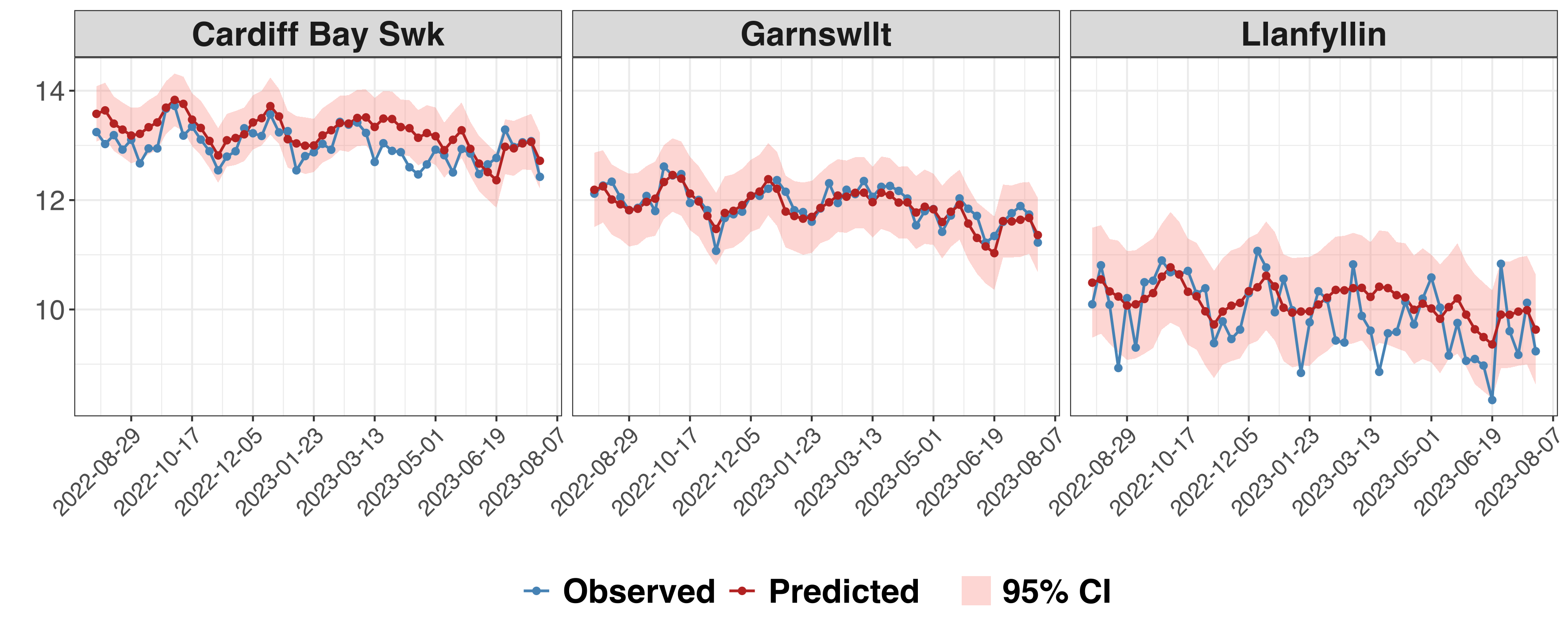}
    \captionof{figure}{Comparison of LGOCV predictions for three sites: Cardiff Bay Swk (biggest site), Llanfyllin (smallest site), and Garnswllt (site with smallest RMSE)}
    \label{fig:lgocv_3sites_scenarioA}
\end{minipage}

\end{center}

\begin{center}

\begin{minipage}{0.4\linewidth}
    \centering
    \includegraphics[width=0.8\linewidth]{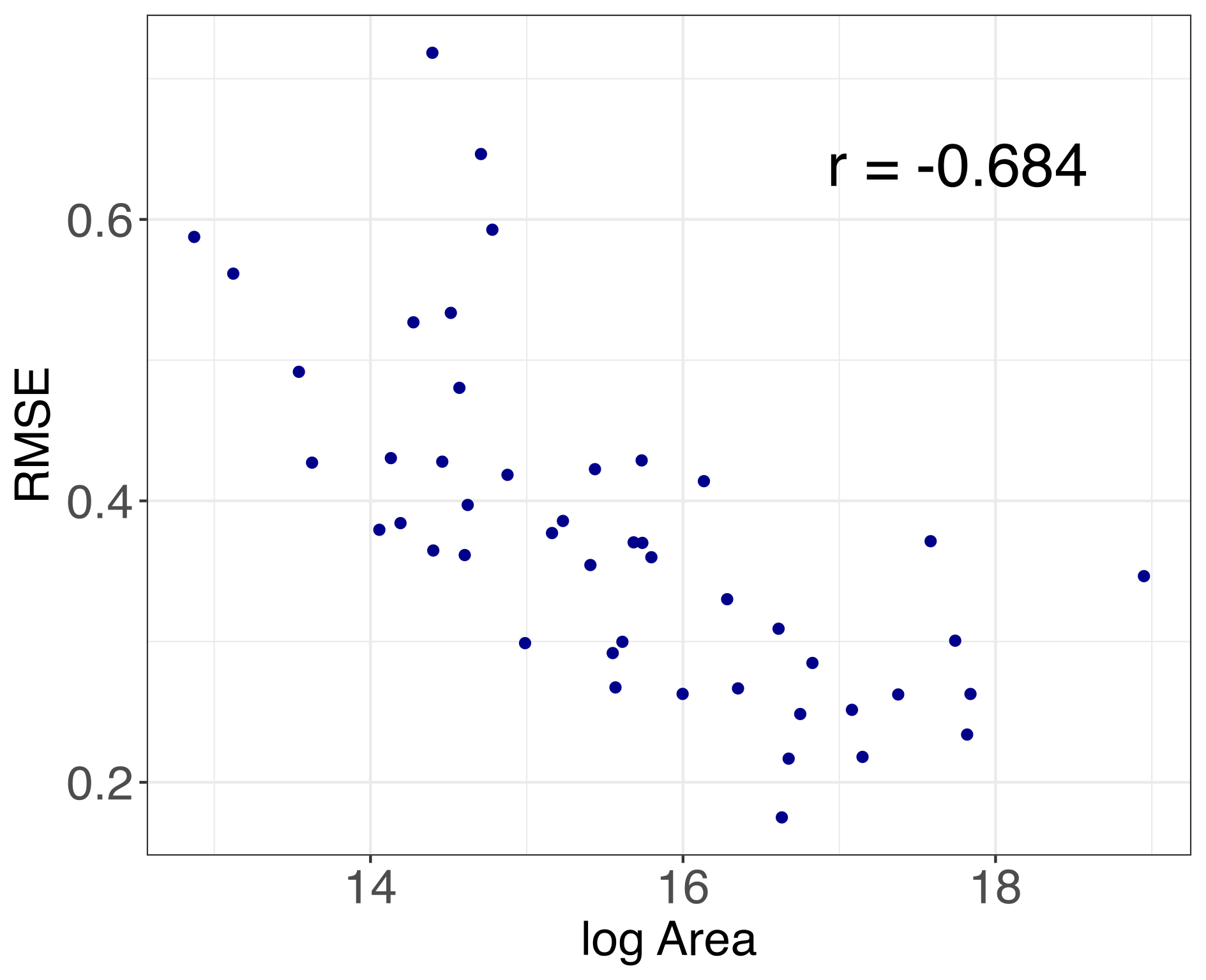}
    
    \vspace{1mm}
    {(a) RMSE}  
\end{minipage}
\hspace{5mm}
\begin{minipage}{0.4\linewidth}
    \centering
    \includegraphics[width=0.8\linewidth]{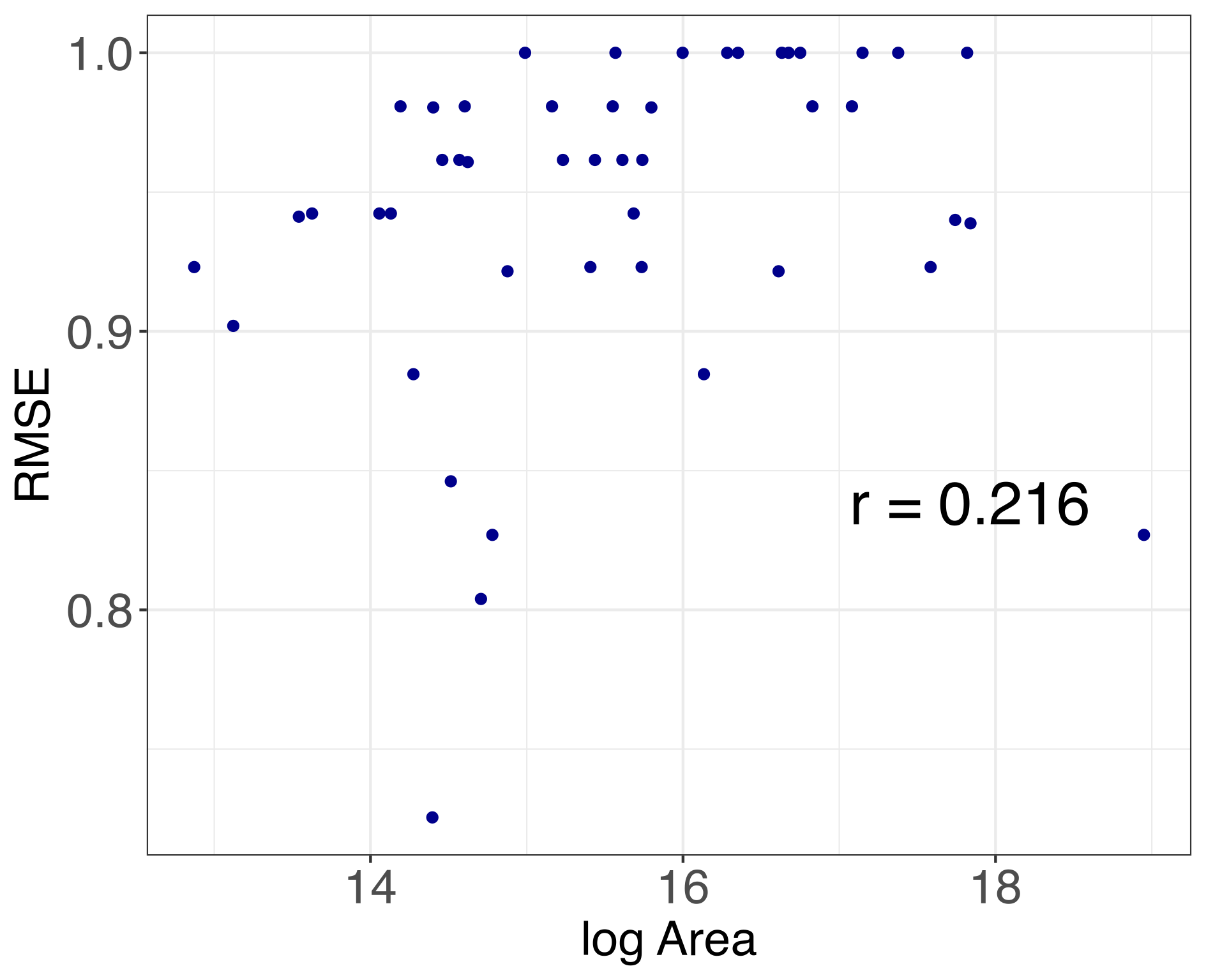}
    
    \vspace{1mm}
    {(b) Coverage}  
\end{minipage}

\end{center}

\captionof{figure}{Scatterplots showing (a) RMSE vs.\ log(Area) and (b) Coverage vs.\ log(Area) from LGOCV predictions. Bigger catchment areas have lower predictive RMSE and higher coverage, except for Cardiff Bay Swk.}
\label{fig:combined_scatter_side_scenA}

\subsection{Predictions at new spatial configurations}\label{subsec:resuts_predictions}

\setlength{\parskip}{0.2em}   
\setlength{\parindent}{15pt} 

A key advantage of the proposed framework is its ability to generate predictions at arbitrary spatial configurations, enabling seamless integration with administrative health data. Figure \ref{fig:illust_swansea} illustrates the spatial predictions for the local authority of Swansea, showing the predicted values of $\log\mathbb{E}\big[ Z_{ijt} \big]$, for week $t=30$ at the nested intersections $b_{ij}$. These are then aggregated up to the LSOA level  and at MSOA level, using Equations \ref{eq:pred_LSOA} and \ref{eq:pred_MSOA_LTLA}, as shown in Figures \ref{fig:illust_swansea}b and \ref{fig:illust_swansea}c, respectively. 

\begin{center}

\begin{minipage}{0.33\linewidth}
    \centering
    \includegraphics[width=0.95\linewidth]{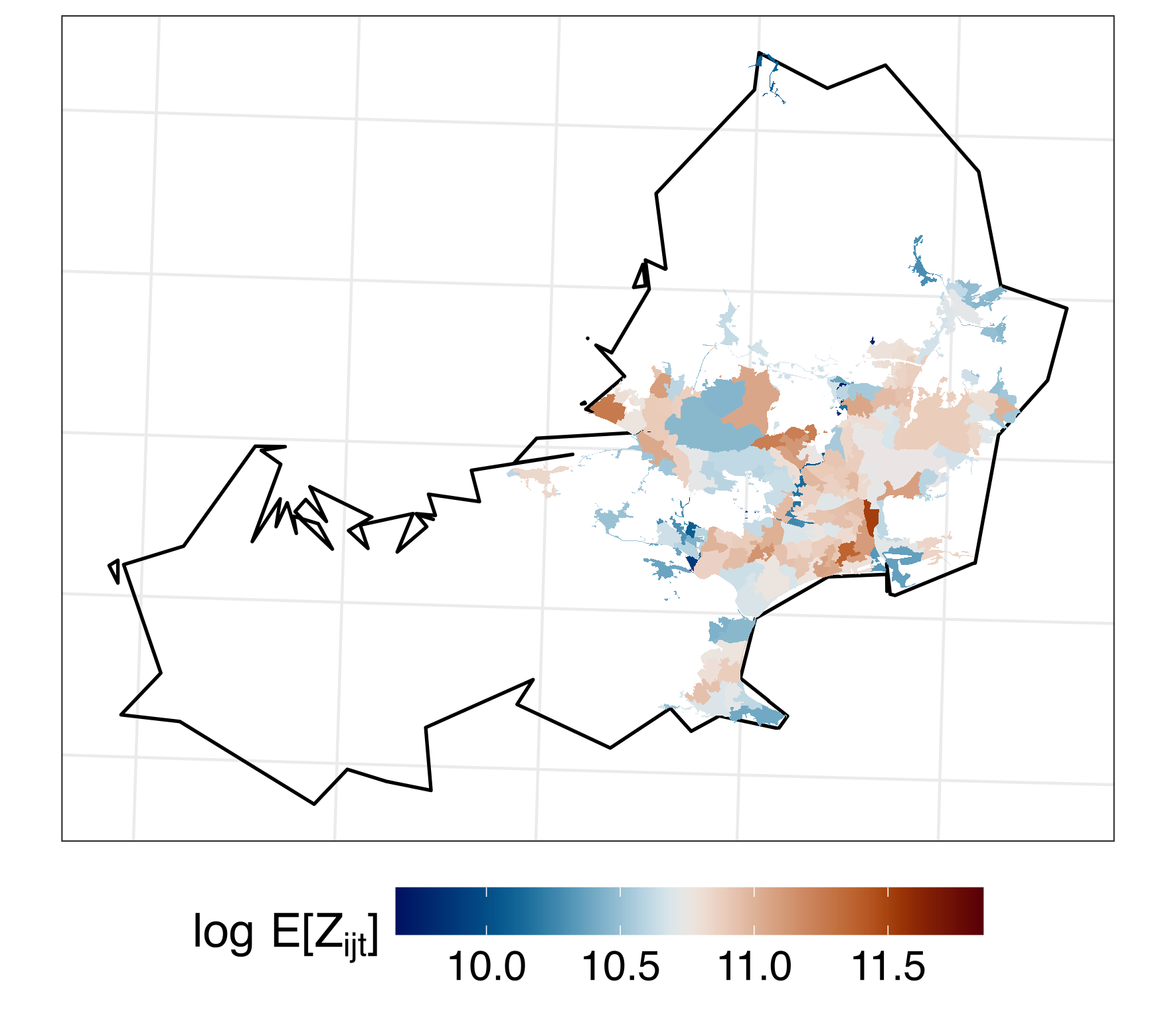}
    
    \vspace{1mm}
    {(a) $\log\mathbb{E}\big[ Z_{ij,30} \big]$}  
\end{minipage}
\begin{minipage}{0.33\linewidth}
    \centering
    \includegraphics[width=0.95\linewidth]{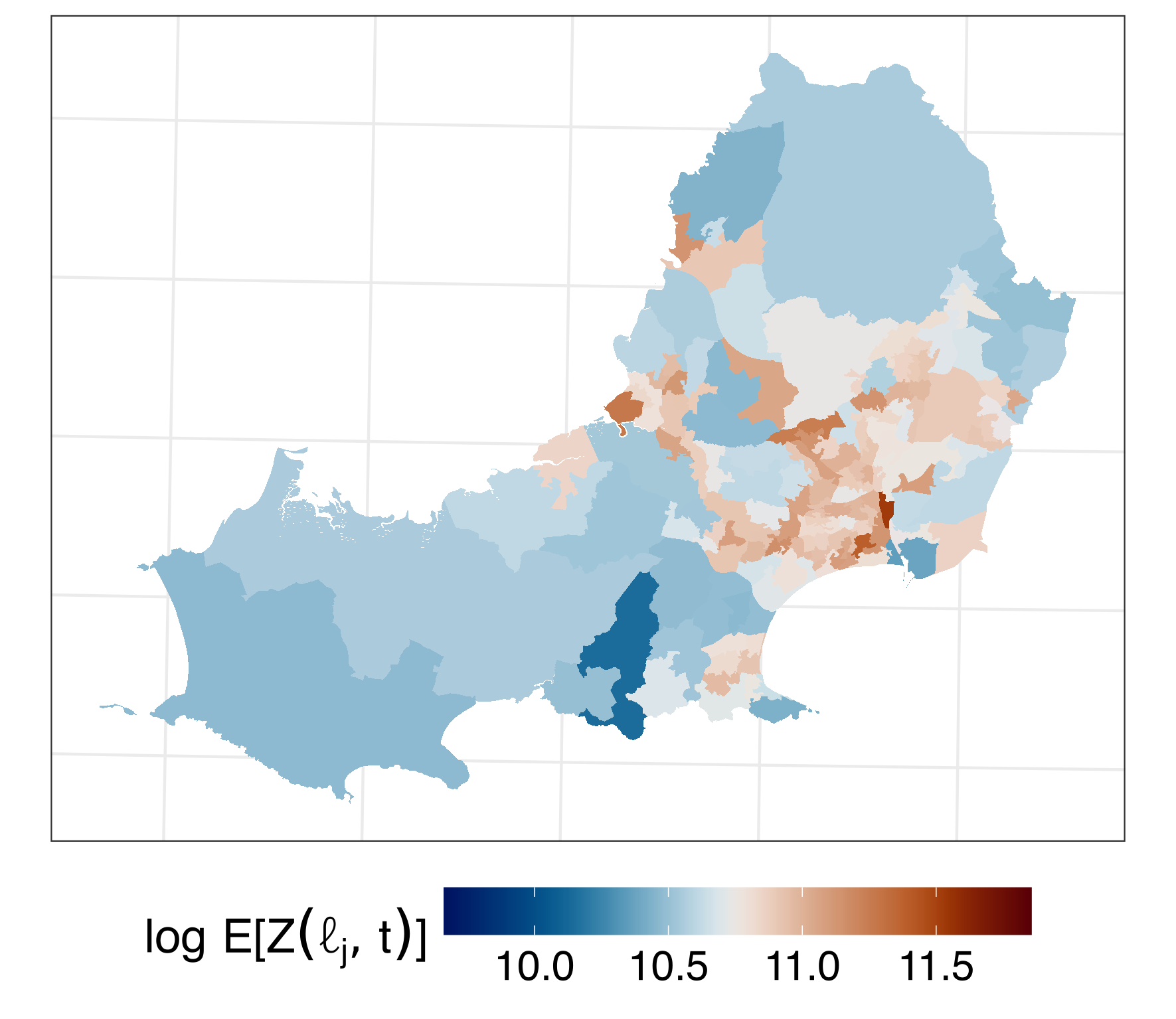}
    
    \vspace{1mm}
    {(b) $\log\mathbb{E}\big[ Z(\text{LSOA}_x,30) \big]$}  
\end{minipage}
\begin{minipage}{0.33\linewidth}
    \centering
    \includegraphics[width=0.95\linewidth]{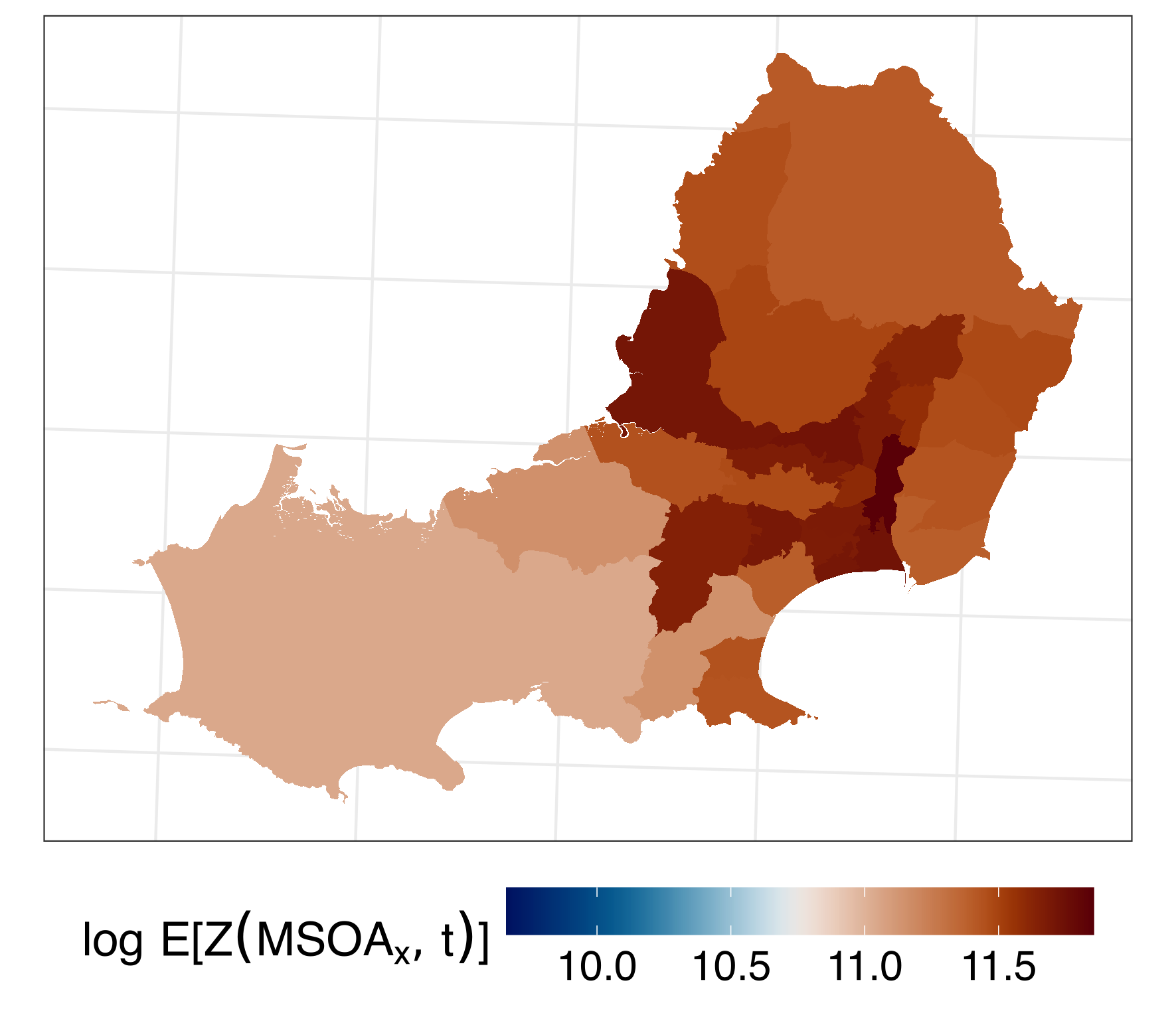}
    
    \vspace{1mm}
    {(b) $\log\mathbb{E}\big[ Z(\text{MSOA}_x,30) \big]$}  
\end{minipage}
\captionof{figure}{Illustration of doing spatial predictions at various spatial boundaries inside the local authority of Swansea: (a) $b_{ij}$-level, (b) LSOAs, (c) MSOAs} \label{fig:illust_swansea}

\end{center}

\begin{center}
    
\begin{minipage}{0.9\linewidth}
    \centering
    \includegraphics[width=\linewidth]{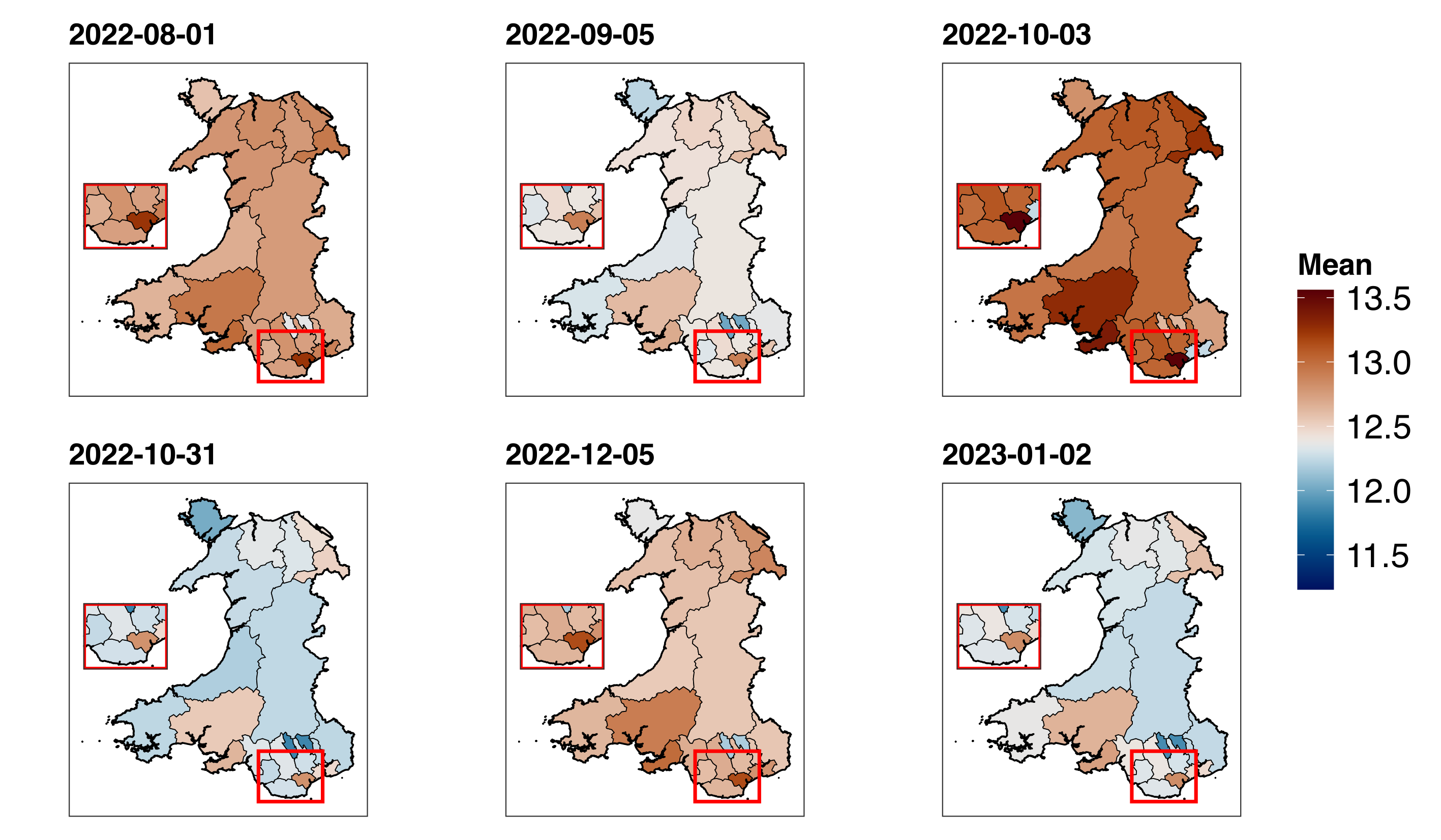}
\end{minipage}
\end{center}
\captionof{figure}{Estimated $\log\mathbb{E}\big[(\text{LTLA}_y,t)\big]$, log (base 10) expected number of gene copies for each LTLA, for weeks from Aug 2022 to Jan 2023.}
\label{fig:pred_mean_LTLA}

\vspace{3mm}

Figure \ref{fig:pred_mean_LTLA} shows the estimated log expected number of gene copies for the 22 LTLAs in Wales, $\log \mathbb{E}\big[\text{LTLA}_{y,t}\big]$, for the first week of each month from August 2022 to January 2023. These estimates were obtained by generating posterior samples from the latent field $\begin{pmatrix} \beta_0^* & \bm{\beta} & \bm{\xi} \end{pmatrix}^\intercal$, evaluating Equations \eqref{eq:pred_LSOA} and \eqref{eq:pred_MSOA_LTLA} for each posterior sample, and then summarising the resulting LTLA-level predictions using their posterior means and standard deviations. Clear spatial and temporal structure is evident.
The Cardiff area (highlighted in red) consistently shows the highest predicted number of gene copies across all time points, reflecting its large population and high urban land cover. A general increase in predicted viral load is observed 
in early October and December 2022, consistent with the seasonal rise in SARS-CoV-2 transmission during the winter period. The corresponding posterior SDs of $\log\mathbb{E}\big[(\text{LTLA}_y,t)\big]$ are shown in Figure \ref{fig:pred_sd_LTLA}. The estimated log expected number of gene copies for each LSOA and MSOA for the same weeks are shown in Figures \ref{fig:pred_mean_LSOA} and  \ref{fig:pred_mean_MSOA}, respectively, in the Appendix.

   \begin{center}
    
\begin{minipage}{0.9\linewidth}
    \centering
    \includegraphics[width=\linewidth]{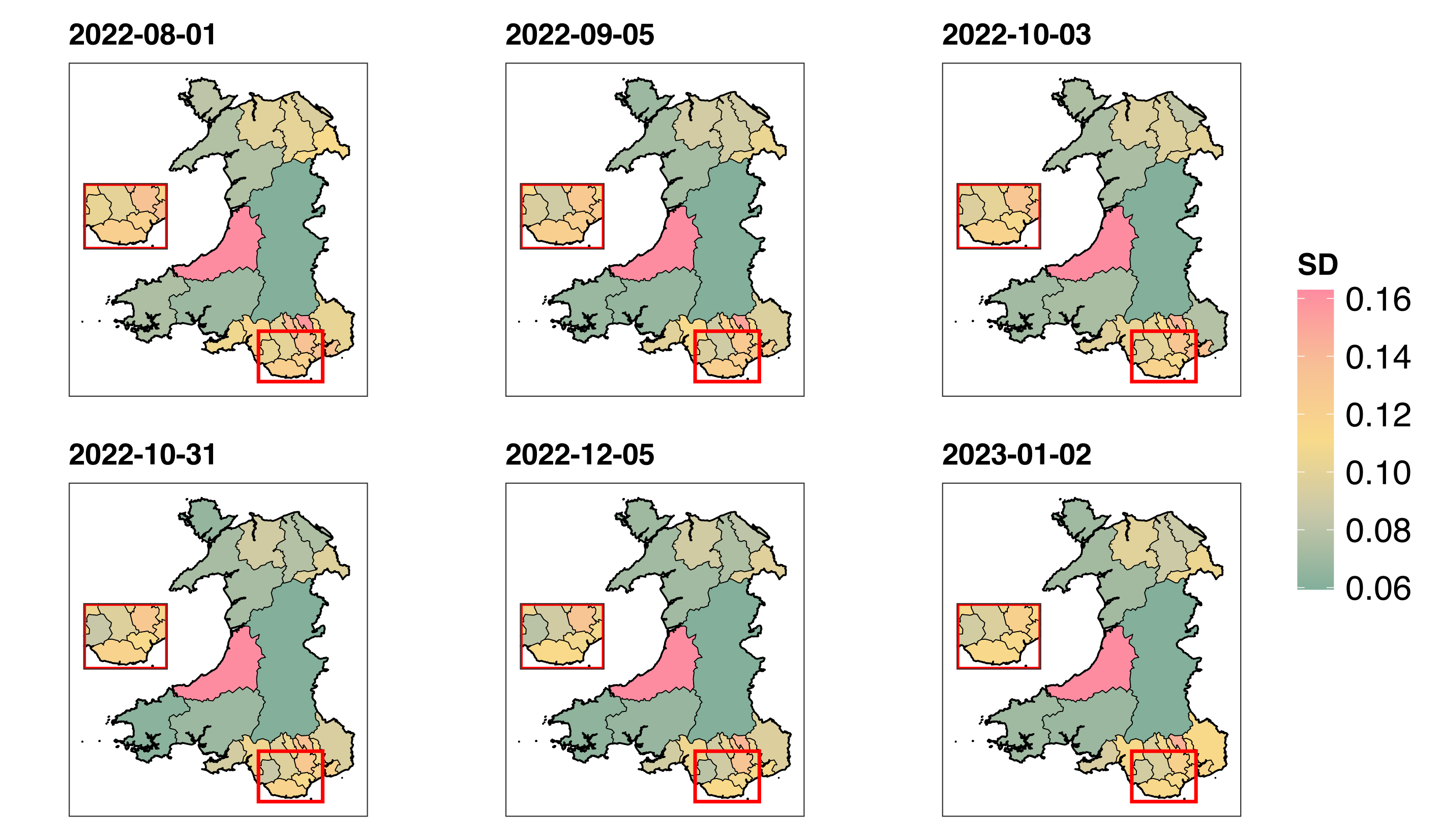}
\end{minipage}

\end{center}

\captionof{figure}{Estimated posterior SD of $\log\mathbb{E}\big[(\text{LTLA}_y,t)\big]$, log (base 10) expected number of gene copies for each LTLA, for weeks from Aug 2022 to Jan 2023.}
\label{fig:pred_sd_LTLA}

\subsection{Association between wastewater signal and PCR positivity rates}\label{subsec:link_ww_pcr}

\setlength{\parskip}{0.2em}   
\setlength{\parindent}{15pt} 

To assess the practical value of the model predictions, we examine the association between the predicted wastewater signal and COVID-19 PCR positivity rates at the LTLA level. An important consideration is that the PCR positivity rate is obtained from laboratory information management systems (LIMS) as the proportion of tests returning a positive result, which reflects a combination of
underlying infection prevalence, testing behaviour, eligibility criteria, healthcare access, and changes in public health policy over time. For this reason, the observed associations between wastewater signals and PCR positivity rates may not reflect the relationship between wastewater signals and the true disease incidence or infection rate. 

Figure \ref{fig:ww_vs_pcr_timeplot} shows time series plots of weekly log-transformed (base $e$) total gene copies and PCR positivity rate for each of the 22 LTLAs. The two signals show broadly similar temporal trends across most LTLAs, with peaks in the wastewater signal generally coinciding with peaks in PCR positivity. Figure \ref{fig:ww_vs_pcr_corrs} presents the corresponding Pearson correlation coefficients. Moderately strong and statistically significant correlations are observed for 21 of the 22 LTLAs, with correlation coefficients ranging from 0.25 (Isle of Anglesey) to 0.62 (Torfaen). The exception is Bridgend (0.15), for which no statistically significant association is detected. This may reflect local differences in catchment characteristics, testing behaviour, or data quality at this site, and warrants further investigation.

The strength of association varies across LTLAs, 
which could arise from differences in population size, wastewater catchment characteristics, testing behaviour, and variability in viral shedding patterns. Overall, these exploratory results support the value of wastewater surveillance as a complementary tool for monitoring community transmission of SARC-CoV-2, particularly in settings where clinical testing data are incomplete or delayed.


\begin{center}
    \includegraphics[width=0.95\linewidth]{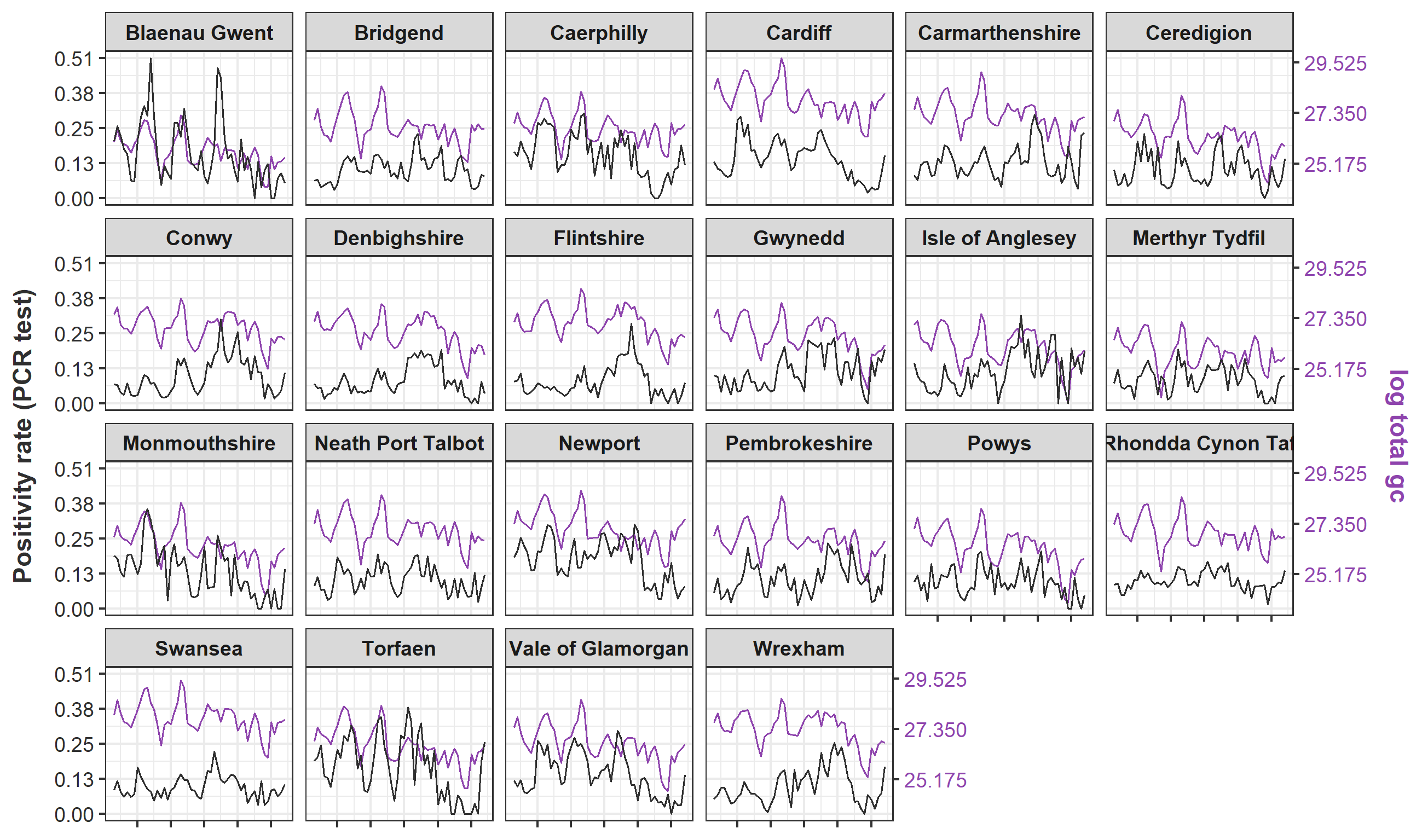}
    \captionof{figure}{Time series plot of weekly log total gene copies of SARS-CoV-19 and PCR positivity rate for each LTLA}
    \label{fig:ww_vs_pcr_timeplot}
\end{center}

\begin{center}
    \includegraphics[width=0.95\linewidth]{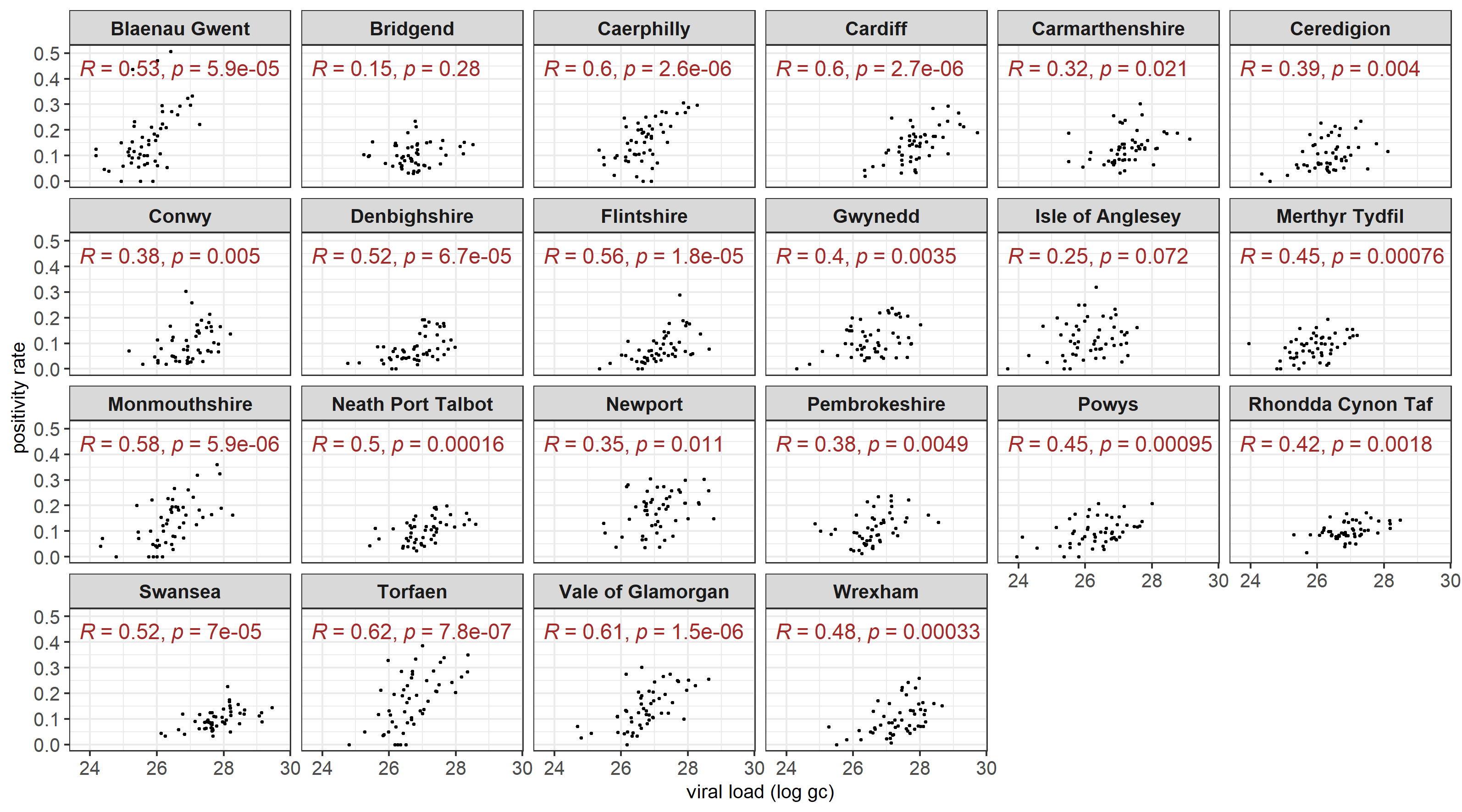}
    \captionof{figure}{Correlation between weekly log total gene copies of SARS-CoV-19 and PCR positivity rate for each LTLA}
    \label{fig:ww_vs_pcr_corrs}
\end{center}

\section{Conclusion}\label{sec:discussion}

We have proposed a flexible approach to
the analysis of wastewater virus concentrations data that combines a spatially continuous latent process and an aggregation of latent Gaussian outcomes over irregular catchment sub-areas. The approach enables inference at multiple spatial scales while accounting for the complex geometry of wastewater catchments and considering the physical motivation of the observed viral load as an aggregation of values from contributing nested areas. Results from leave-group-out crossvalidation (LGOCV) indicate that the model achieves good predictive performance, while the observed association between predicted wastewater signals and PCR positivity rates demonstrates the practical value of the generated estimates. However, weaker predictive performance in small catchments warrants further investigation.

A major advantage of the proposed framework is that it separates the fine-scale latent model and the coarse-scale sampling model. Unlike conventional approaches that perform pre-processing steps to match fine-scale covariates and the block configuration of the outcome data, making them prone to the Modifiable Areal Unit Problem (MAUP) \citep{manley2021scale}, the proposed methodology provides a seamless connection between the latent spatially continuous model and the sampling model that takes into account the aggregation of the 
spatially continuous latent process. This framework enables prediction at arbitrary spatial (areal) configurations, facilitating direct integration with routinely collected health data. The approach is an example of a block aggregation model, as proposed in \cite{villejo2026spatially}, but extended to
incorporate a sampling distribution that is not closed under aggregation, such as the log Gaussian distribution.

A long-term objective is to use wastewater signals as an early warning indicator for health outcomes such as hospital admissions and disease incidence. While the proposed framework provides a flexible approach for generating such signals, its computational complexity may pose challenges for routine implementation. In particular, regular model fitting and forecasting would require reliable computational infrastructure, especially when near real-time surveillance is required. To facilitate routine deployment, inference was performed using the linearised Integrated Nested Laplace Approximation (INLA), implemented using the \texttt{inlabru} library, which provides accurate approximate Bayesian inference at a substantially lower computational cost than simulation-based methods such as Markov chain Monte Carlo (MCMC). 
Further reductions in the computational burden, albeit at the cost of some loss of predictive precision, could be obtained by using a smaller network of strategically selected sentinel sampling sites over the region of interest.


\section*{Funding}
The work has been funded by the Medical Research Council (MRC), award UKRI078: `Incorporating wastewater-based epidemiology into a real-time, multiplex public health surveillance system'. MB also acknowledges partial support from the MRC Centre for Environment and Health, funded by the UK Medical Research Council, Grant number: MR/L01341X/1. MB, EW,  and SV acknowledge infrastructure support for the Department of Epidemiology and Biostatistics provided by the NIHR Imperial Biomedical Research Centre (BRC).

\printcredits

\bibliographystyle{cas-model2-names}

\bibliography{cas-refs}


\section*{Appendix}

\begin{appendix}
    \section{Approximate distribution of $S_{it}$}

    Suppose we have nested grids $\big\{b_{i1}, \ldots, b_{iJ_i} \big\}$ that form a partition of an area $C_i$. We denote by $Z_{ijt}$ as the number of gene copies at $b_{ij}$  for time $t$, $j=1,\ldots,J_i$ and $t=1,\ldots,T$. We assume
\begin{equation}
    Z_{ijt} \sim \log \text{Normal}\big(\mu_{ijt},\sigma^2_Z\big).
\end{equation}

    We have the following moments of $Z_{ijt}$:
\begin{equation*}
    \begin{aligned}
        &\mathbb{E}(Z_{ijt}) = \exp\Big\{\mu_{ijt}+\dfrac{1}{2}\sigma^2_Z\Big\} \equiv m_{ijt} \\
        &\mathbb{V}(Z_{ijt}) = \Big(\exp\big\{\sigma^2_Z\big\}-1\Big)\Big(\exp\big\{ 2\mu_{ijt} + \sigma^2_Z\big\}\Big) \equiv v_{ijt}. 
    \end{aligned}
\end{equation*}

    We observe an aggregated value, say $S_{it}=\sum_{j=1}^{J_i} Z_{ijt}$. Suppose that we assume $S_{it} \approx \log\text{Normal}\Big(\mu_{S_{it}}, \sigma^2_{S_{it}}\Big)$.

    This implies that:
\begin{equation*}
    \begin{aligned}
        &\mathbb{E}(S_{it}) = \exp\Big\{\mu_{S_{it}}+\dfrac{1}{2}\sigma^2_{S_{it}}\Big\} \\
        &\mathbb{V}(S_{it}) = \Big(\exp\big\{\sigma^2_{S_{it}}\big\}-1\Big)\Big(\exp\big\{ 2\mu_{S_{it}} + \sigma^2_{S_{it}}\big\}\Big) 
    \end{aligned}
\end{equation*}

The mean and variance of $S_{it}$ are given by
 \begin{equation}
    \begin{aligned}
        &\mathbb{E}(S_{it}) = \sum_{j=1}^{J_i} \mathbb{E}(Z_{ijt}) = \sum_{j=1}^{J_i} m_{ijt} \\
        &\mathbb{V}(S_{it}) = \sum_{j=1}^{J_i} \mathbb{V}(Z_{ijt}) +  2\sum_{j<k}\text{Cov}(Z_{ijt},Z_{ikt}) = \sum_{j=1}^{J_i} v_{ijt} +  2\sum_{j<k}\text{Cov}(Z_{ijt},Z_{ikt})
    \end{aligned}
\end{equation}
The covariance term is given by
\begin{equation}
    \begin{aligned}
        \text{Cov}(Z_{ijt},Z_{ikt}) &=\exp\big\{ \mu_{ijt} + \mu_{ikt} + \sigma^2_Z \big\}\times\Big[\exp\big\{c_{ijk,t}\big\}-1\Big]\\
        &=m_{ijt}m_{ikt}\times\Big[\exp\big\{c_{ijk,t}\big\}-1\Big]
    \end{aligned}
\end{equation}
where $c_{ijk,t} = \text{Cov}(\log Z_{ijt},\log Z_{ikt})$.

\vspace{5mm}

Matching the moments, we have
\begin{equation}\label{eq:moment1}
    \sum_{j=1}^{J_i} m_{ijt} \equiv \exp\Big\{\mu_{S_{it}}+\dfrac{1}{2}\sigma^2_{S_{it}}\Big\}, \;\;\; \text{and}
\end{equation}
\begin{equation}\label{eq:moment2}
\sum_{j=1}^{J_i} v_{ijt} +  2\sum_{j<k}\text{Cov}(Z_{ijt},Z_{ijk}) \equiv \Big(\exp\big\{\sigma^2_{S_{it}}\big\}-1\Big)\Big(\exp\big\{ 2\mu_{S_{it}} + \sigma^2_{S_{it}}\big\}\Big).
\end{equation}

Equation \eqref{eq:moment1} yields
\begin{equation*}
    \mu_{S_{it}} = \log \sum_{j=1}^{J_i} m_{ijt} - \dfrac{1}{2}\sigma^2_{S_{it}}.
\end{equation*}
Plugging in the form of $\mu_{S_{it}}$ to Equation \eqref{eq:moment2} yields
\begin{equation*}
    \begin{aligned}
        \sum_{j=1}^{J_i} v_{ijt} +  2\sum_{j<k}\text{Cov}(Z_{ijt},Z_{ijk}) &\equiv \Big(\exp\big\{\sigma^2_{S_{it}}\big\}-1\Big)\Big(\exp\Big\{ 2\Big[\log \sum_{j=1}^{J_i} m_{ijt} - \dfrac{1}{2}\sigma^2_{S_{it}}\Big] + \sigma^2_{S_{it}}\Big\}\Big)\\
        &=\Big(\exp\big\{\sigma^2_{S_{it}}\big\}-1\Big)\exp\Big\{ 2\log\sum_{j=1}^{J_i} m_{ijt}\Big\}.
    \end{aligned}
\end{equation*}
Solving for $\sigma^2_{S_{it}}$, we get $\exp\big\{\sigma^2_{S_{it}}\big\}-1 = \dfrac{\sum_{j=1}^{J_i} v_{ijt} +  2\sum_{j<k}\text{Cov}(Z_{ijt},Z_{ikt})}{\exp\Big\{ 2\log\sum_{j=1}^{J_i} m_{ijt}\Big\}}$, which gives
\begin{equation*}
    \sigma^2_{S_{it}} = \log \Bigg( \dfrac{\sum_{j=1}^{J_i} v_{ijt} +  2\sum_{j<k}\text{Cov}(Z_{ijt},Z_{ikt})}{\exp\Big\{ 2\log\sum_{j=1}^{J_i} m_{ijt}\Big\}} + 1\Bigg) = \log \Bigg( \dfrac{\sum_{j=1}^{J_i} v_{ijt} +  2\sum_{j<k}\text{Cov}(Z_{ijt},Z_{ikt})}{\big( \sum_{j=1}^{J_i}m_{ijt} \big)^2} + 1\Bigg).
\end{equation*}

\vspace{5mm}

\begin{center}

\begin{minipage}{0.8\linewidth}
    \centering
    \includegraphics[width=\linewidth]{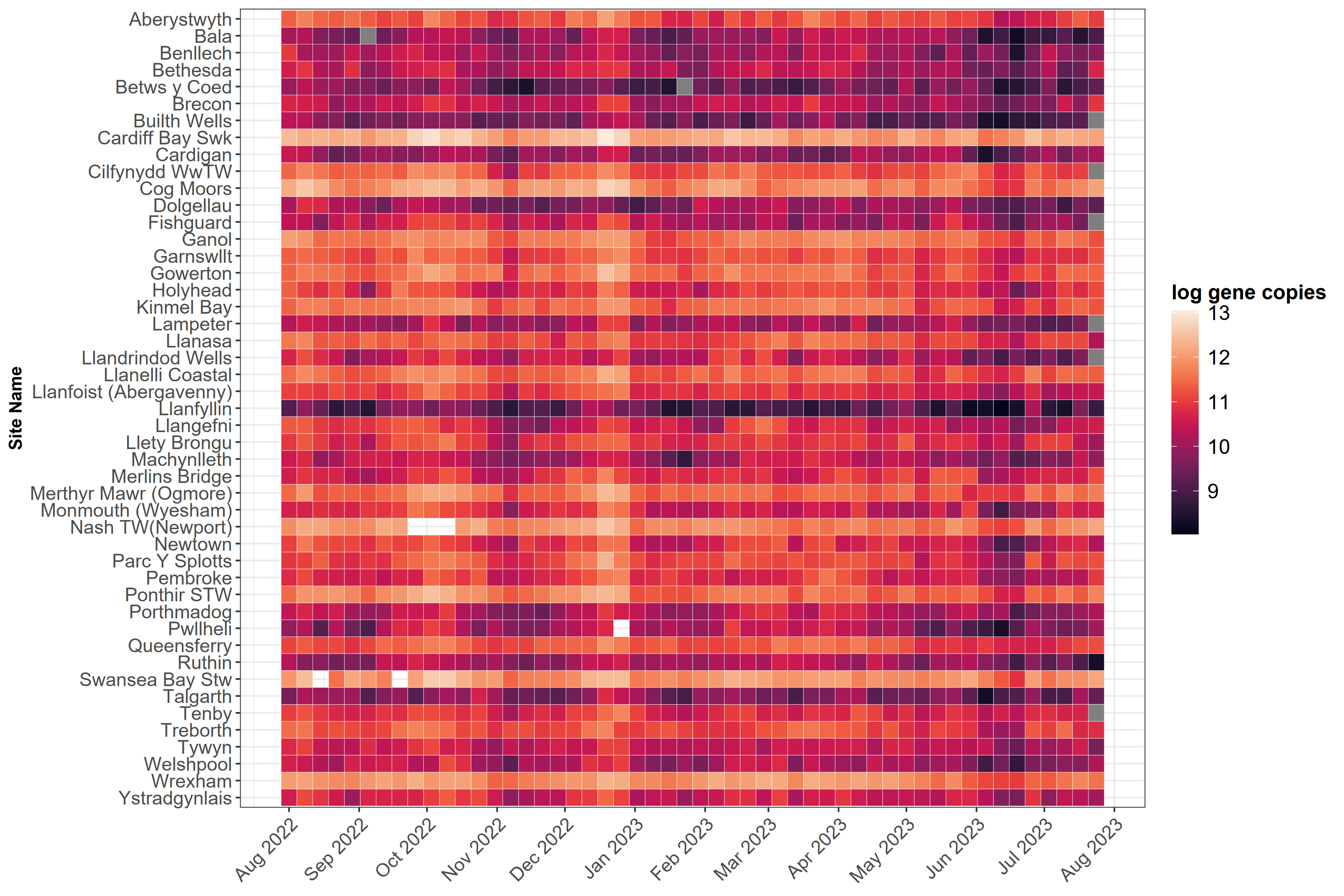}
\end{minipage}

\end{center}

\captionof{figure}{Heatmap of weekly log viral load (flow-normalized concentrations) of SARS-CoV-2N1 for each catchment site from August 2022 to July 2023}
\label{fig:heatmap}

\begin{center}

\begin{minipage}{0.8\linewidth}
    \centering
    \includegraphics[width=\linewidth]{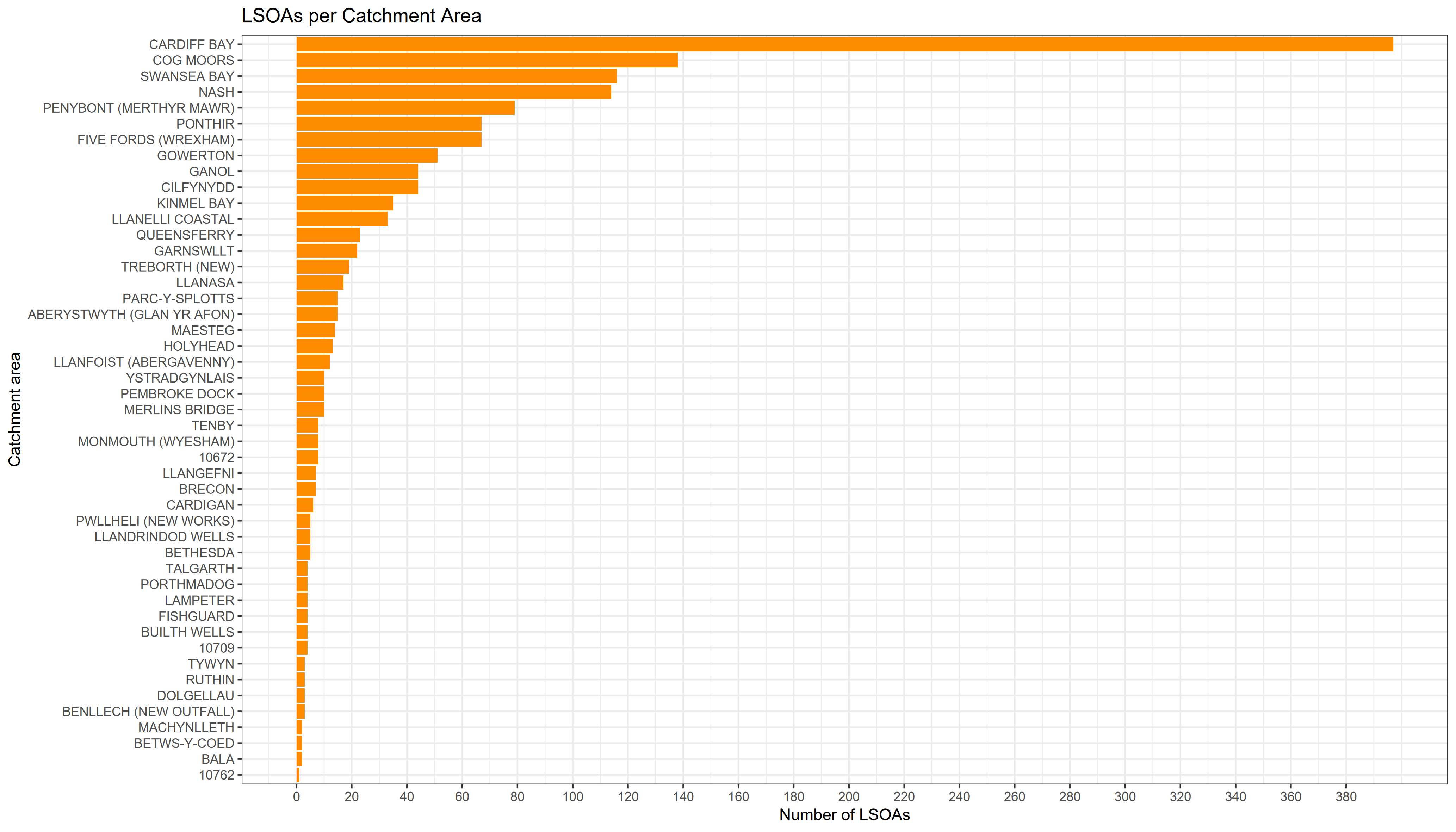}
\end{minipage}

\end{center}

\captionof{figure}{Number of LSOAs served by each catchment area}
\label{fig:lsoas_per_catchment}

\begin{center}

\begin{minipage}{0.9\linewidth}
    \centering
    \includegraphics[width=\linewidth]{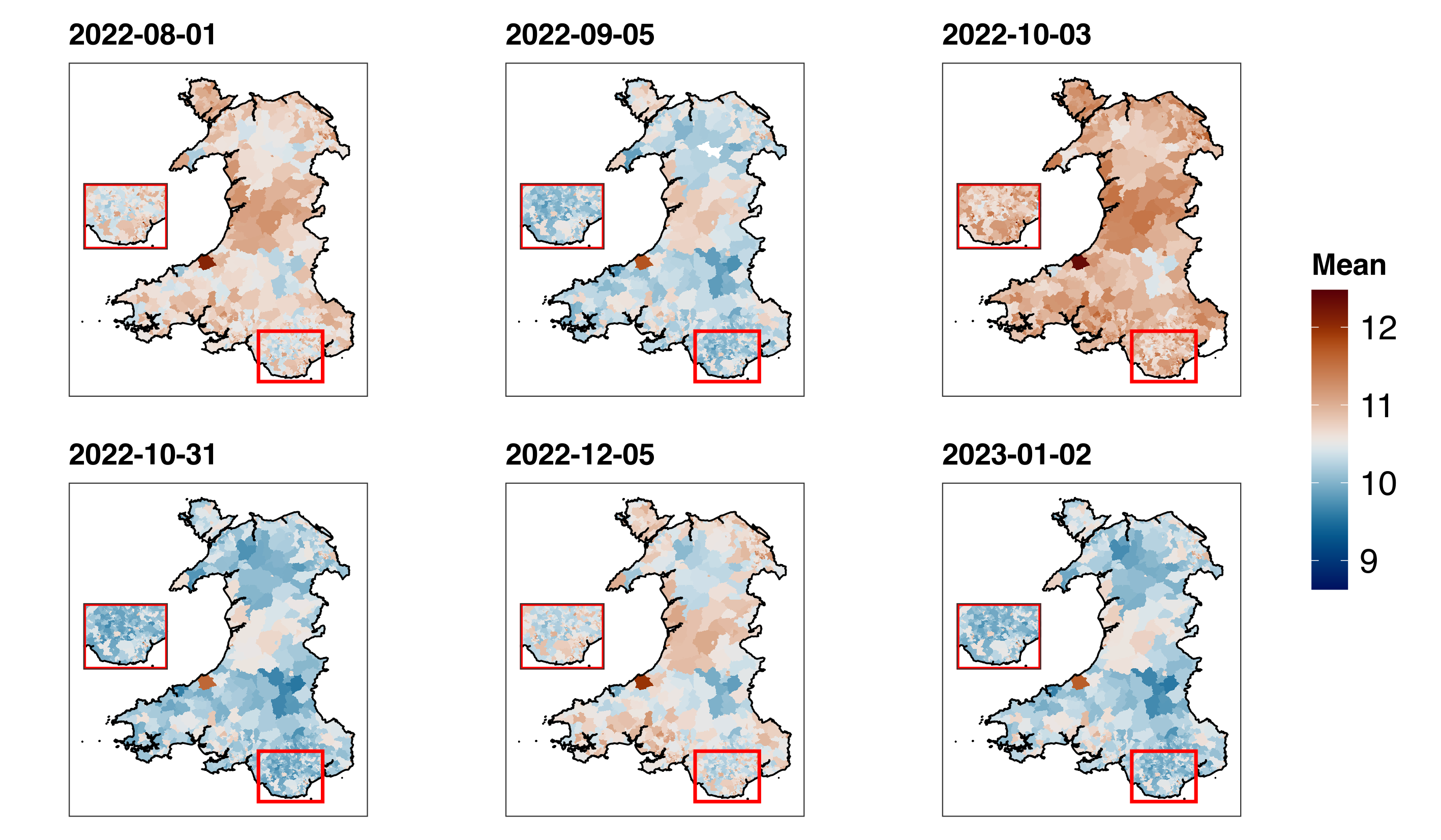}
\end{minipage}

\end{center}

\captionof{figure}{Predicted log (base 10) number of gene copies for each LSOA, for weeks from Aug 2022 to Jan 2023.}
\label{fig:pred_mean_LSOA}

\begin{center}

\begin{minipage}{0.9\linewidth}
    \centering
    \includegraphics[width=\linewidth]{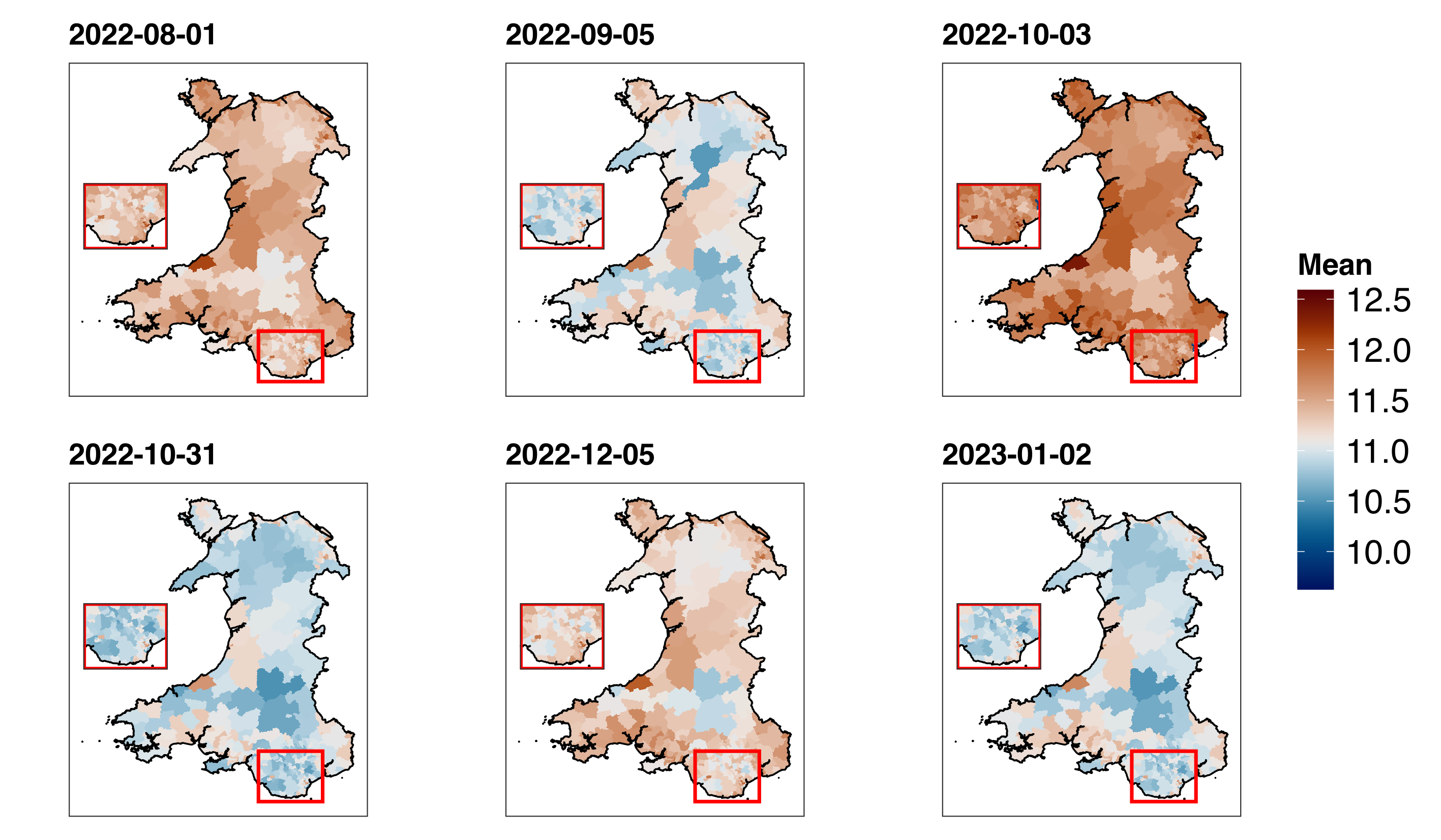}
\end{minipage}

\end{center}
\captionof{figure}{Predicted log (base 10) number of gene copies for each MSOA, for weeks from Aug 2022 to Jan 2023.}
\label{fig:pred_mean_MSOA}

\begin{landscape}

\begin{figure}[p]
    \centering
    \includegraphics[width=0.99\linewidth]{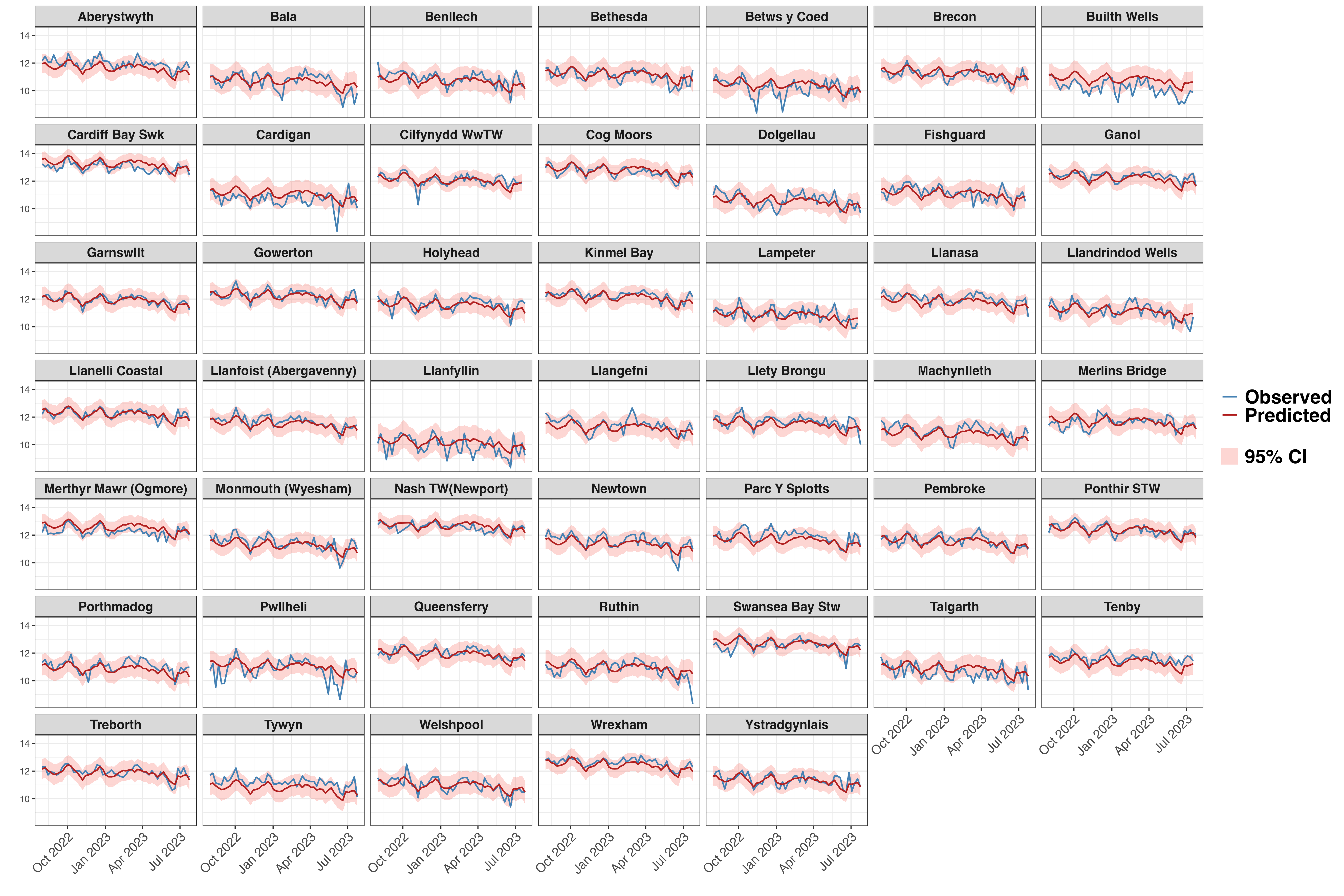}
    \caption{LGOCV predictions for all catchment areas}
    \label{fig:CVplotsall}
\end{figure}
    
\end{landscape}


\end{appendix}



\end{document}